\documentclass[12pt]{article}
\usepackage{color,amssymb,amsmath,youngtab}
\usepackage{url}

 \csname
@addtoreset\endcsname{equation}{section}

\def\a{\alpha} \def\ad{\dot{\a}} 
\def\ua{\underline{\alpha}}
\def\ub{\underline{\phantom{\alpha}}\!\!\!\beta}

\def\b{\beta}  \def\bd{\dot{\b}} 
\def\c{\gamma} \def\cd{\dot{\c}}

\def\d{\delta} \def\dd{\dot{\d}}

\def\e{\epsilon}

\def\k{\kappa}
\def\l{\lambda}
\def\L{\Lambda}
\def\m{\mu}

\def\s{\sigma}

\def\t{\tau}
\def\x{\xi}
\def\y{\eta}

\def\O{\Omega}
\def\o{\omega}

\def\th{\theta}

\def\cN{{\cal N}}

\def\yb{{\bar y}}
\def\zb{{\bar z}}

\def\del{\partial}

\let\la=\label

\def\nn{\nonumber}
\newcommand{\eq}[1]{(\ref{#1})}
\newcommand{\ns}[1]{{\normalsize #1}}

\newcommand{\w}[1]{\\[0.#1cm]}
 \def\det{{\rm det\,}}
\def\be{\begin{equation}}
\def\ee{\end{equation}}
\def\bea{\begin{eqnarray}}
\def\eea{\end{eqnarray}}
\def\ba{\begin{array}}
\def\ea{\end{array}}
\def\se{\;\;=\;\;}

\def\ft#1#2{{\textstyle{{\scriptstyle #1}
\over {\scriptstyle #2}}}}

\def\ed{\end{document}}

\def\Real{{\mathbb R}}
\def\Comp{{\mathbb C}}

\def\sb{\bar\sigma}

\def\AS{{\tiny\begin{tabular}{|c|}
\hline
\\
\hline
\\
\hline
\end{tabular}}}

\def\SY{
{\tiny\begin{tabular}{|l|l|}
\hline
&\\
\hline
\end{tabular}}}

\def\MAT{
{\tiny\begin{tabular}{|l|l|}
\hline
\!\!$\times$\!\!&\\
\hline
\end{tabular}}}

\begin{document}

\begin{titlepage}

\setcounter{page}{1}

\begin{center}

\hfill MIFPA-12-31

\vskip 1cm


{\Large \bf  Supersymmetric Higher Spin Theories}


\vspace{12pt}

Ergin Sezgin\,$^1$ and Per Sundell\,$^{2,}$\footnote[3]{Ulysse Incentive Grant for Mobility in 
Scientific Research, F.R.S.-FNRS} \\

\vskip 25pt

{\em $^1$ \hskip -.1truecm George and Cynthia Woods Mitchell Institute for Fundamental Physics 
and Astronomy, Texas A\& M University, College Station,
TX 77843, USA}

{email: {\tt sezgin@tamu.edu}}

\vskip 15pt

{\em $^2$ \hskip -.1truecm Service de M\`ecanique et Gravitation, Universit\'e de Mons \\
20, Place du Parc, B7000 Mons, Belgium\vskip 5pt }

{email: {\tt per.sundell@umons.ac.be}} \\

\end{center}

\vskip 1cm

\begin{center} {\bf ABSTRACT}\\[3ex]
\end{center}

We revisit the higher spin extensions of the anti de Sitter algebra in four dimensions that incorporate internal symmetries and admit representations that contain fermions, classified long ago by Konstein and Vasiliev. We construct the $dS_4$, Euclidean and Kleinian version of these algebras, as well as the corresponding fully nonlinear Vasiliev type higher spin theories, in which the reality conditions we impose on the master fields play a crucial role. The  ${\cal N}=2$ supersymmetric higher spin theory in $dS_4$, on which we elaborate further, is included in this class of models. A subset of Konstein-Vasiliev algebras are the higher spin extensions of the $AdS_4$ superalgebras $osp(4|{\cal N})$ for ${\cal N}=1,2,4$ mod $4$ and can be realized using fermionic oscillators. We tensor the higher superalgebras of the latter kind with appropriate internal symmetry groups and show that the ${\cal N}=3$ mod $4$ higher spin algebras are isomorphic to those with ${\cal N}=4$ mod $4$.  We describe the fully nonlinear higher spin theories based on these algebras as well, and we elaborate further on the ${\cal N}=6$  supersymmetric theory, providing two equivalent descriptions one of which exhibits manifestly its relation to the ${\cal N}=8$ supersymmetric higher spin theory.

\end{titlepage}

\newpage

\tableofcontents

\pagebreak


\section{Introduction}


Minimal higher spin (HS) gravity in four dimensional spacetime describes a generalization of Einstein gravity by inclusion of a real scalar field and infinite set of even HS fields, each occurring once, in terms of nonlinear and consistent field equations proposed by Vasiliev in 1990 \cite{Vasiliev:1990en}. For reviews, see \cite{Vasiliev:1995dn,Vasiliev:1999ba,Bekaert:2005vh}.  In recent years a strong motivation for exploring HS theories has been their  holographic descriptions by means of remarkably simple boundary CFTs\cite{Sezgin:2002rt,Klebanov:2002ja}, the weak-weak coupling nature of these descriptions, and the prospects of unraveling a highly symmetric phase of string theory. 

While most investigations in HS theories have been in the context of bosonic models in four dimensions, efforts towards understanding the string/M theory origin of these theories are expected to involve supersymmetry. In fact, historically, a connection between massless HS fields and free boundary CFT seems to have been first proposed in 1988 by Bergshoeff, Salam, Sezgin and Tanii in \cite{Bergshoeff:1988jm} where it was shown that the $osp(8|4)$ invariant superconformal field theory \cite{Bergshoeff:1988jx}, which arises in the quantization of the supermembrane in $AdS_4\times S^7$ with its worldvolume wrapped round the boundary of $AdS_4$ \cite{Duff:1987qa,Bergshoeff:1987dh,Blencowe:1987bn,Nicolai:1988ek}, contains in its spectrum an infinite set of massless HS fields. 
These states correspond to the symmetric product of two $OSp(8|4)$ singletons and arrange themselves into an infinite tower of $osp(8|4)$ supermultiplets with maximum spin $s_{\rm max}=2,4,6, .., \infty$, the lowest one being that of gauged $N=8$ supergravity in four dimensions $(4D)$\footnote{The fact that the symmetric product of two $SO(3,2)$ singletons contains an infinite set of massless HS fields was discovered by Flato and Fronsdal \cite{Flato:1978qz}.}. The details of the underlying super HS theory were spelled out much later in \cite{Sezgin:1998gg,Sezgin:1998eh}.

In the context of HS holography, the next appearance of supersymmetric boundary CFTs occurred in 2002 \cite{Sezgin:2002rt} where super HS theories in $D=4,5,7$ were considered.  These dimensions were motivated by $AdS_5\times S^5$ vacuum of Type IIB string, and the $AdS_4\times S^7$  and the$AdS_7\times S^4$ vacua of M theory. The following year, aspects of HS holography in the context of ${\cal N}=1$ super HS theories in $4D$ were treated in \cite{Leigh:2003gk,Sezgin:2003pt}. More recently, an ${\cal N}=6$ supersymmetric HS theory in 4D and its connection with string theory has been investigated in \cite{Chang:2012kt}.

In formulating the super HS theories in their own right, it is natural to start with the determination of a suitable HS superalgebra. Various HS superalgebras in $4D$ were proposed in \cite{Fradkin:1986ka,Vasiliev:1986qx,Fradkin:1987ah}. However, once the importance of the HS spectrum being determined by the two-fold product of the singletons was understood \cite{Sezgin:1998gg,Konstein:1989ij}, it soon became clear that a sensible HS algebra that underlies a consistent description of HS interactions must have generators in one to one correspondence with the spin $s\ge 1$ part of the spectrum that follows from the two-singleton product. An alternative way to state this requirement is that 
the HS algebra must admit a massless unitary representation with the same spectra of spins as predicted by the structure of gauge fields originating from the HS algebra. This requirement was called the ``admissibility condition'' in \cite{Konstein:1989ij}, and  it was shown in \cite{Konshtein:1988yg} that only a subset of the HS algebras of \cite{Fradkin:1986ka,Vasiliev:1986qx,Fradkin:1987ah} were actually admissible. A class of admissible HS superalgebras\footnote{The terminology of ``HS superalgebra'' is used in a wider context in \cite{Konstein:1989ij} to include algebras that contain fermionic generators but not necessarily contain the ordinary spacetime superalgebras as finite dimensional subalgebras.}  in $4D$ were subsequently constructed by Konstein and Vasiliev (KV) \cite{Konstein:1989ij}.
Further HS superalgebras in $4D$ \cite{Engquist:2002vr}, in $5D$ \cite{Sezgin:2001yf} and $7D$ \cite{Sezgin:2002rt}, and generically in $D>4$ dimensions have also been constructed \cite{Vasiliev:2004cm}. The construction of these algebras uses fermionic or vector oscillators, and in some cases they are extended by tensoring with suitable matrix and Clifford algebras. 

Given a HS superalgebra in $D\le 4$, the construction of the corresponding fully nonlinear HS theory proceeds uneventfully alongs the lines of the minimal Vasiliev system. In each case, it is important to impose suitable reality and projection conditions on the master fields, which depend on a set of commuting fermionic oscillators, and to ensure the correct spin-statistics for their component fields. In dimensions $D>4$, however, the situation is different because the certain properties of the fermionic oscillators that hold in $D\le 4$ cease to hold. For example, in $5D$, it is not clear how to construct an appropriate central term in curvature constraints in a Cartan integrable fashion for the simple reason that the fermionic oscillators are 4-component, while in $4D$ this is not a problem owing to the fact that the fermionic oscillators are 2-component in that case. Nonetheless, HS field equations can still be written down at the lineraized level for models based on HS superalgebras in certain cases; see \cite{Sezgin:2001yf} for $D=5$ and \cite{Sezgin:2002rt} for $D=7$. Employing suitable vector oscillators \cite{Vasiliev:2004cp} instead of fermionic oscillators, bosonic HS theories in arbitrary dimensions have been constructed \cite{Vasiliev:2004cm}, but supersymmetric extension of these models is apparently not known. 

The purpose of this paper is to extend in various directions the results that have been obtained so far on fully nonlinear HS theories with fermions and supersymmetry in four dimensions \cite{Konshtein:1988yg,Vasiliev:1992ix,Vasiliev:1992av,Sezgin:1998gg,Sezgin:1998eh,Engquist:2002vr,Sezgin:2002ru,Chang:2012kt}. More specifically,  we shall construct the $dS_4$, Euclidean and Kleinian version of KV algebras, and the corresponding fully nonlinear Vasiliev type HS theories, in which the reality conditions on the master fields play a crucial role. The ${\cal N}=2$ supersymmetric HS theory in $dS_4$, on which we shall elaborate further, is included in this class of models. A subset of Konstein-Vasiliev algebras are the HS extensions of the $AdS_4$ superalgebras $osp(4|{\cal N})$ for ${\cal N}=1,2,4$ mod $4$ and can be realized using fermionic oscillators \cite{Konstein:1989ij}. We shall tensor the HS superalgebras of the latter kind with appropriate internal symmetry groups and show that the resulting HS algebras with ${\cal N}=3$ mod $4$ are isomorphic to those with ${\cal N}=4$ mod $4$.  We shall provide the fully nonlinear HS theories based on these algebras as well. We shall also exhibit the  truncations of a model with ${\cal N}$ supersymmetry to that with ${\cal N}-2$, and in particular obtain an ${\cal N}=6$ HS theory from an ${\cal N}=8$ HS theory, which provides an alternative description to that of \cite{Chang:2012kt}, and makes manifest its relation to the ${\cal N}=8$ HS theory.


\section{The Bosonic Models in Four Dimensions}


In this section, we review the structure of Vasiliev's equations in the case of bosonic models in $AdS_4$, including their $\widehat\Theta$-deformations, HS gauge symmetries and the minimal projections, in particular defining the Type A and Type B models. We shall then review their formulation in $dS_4$ as well spacetimes with Euclidean, Kleinian signatures.


\subsection{Vasiliev Equations in $AdS_4$}


The four dimensional minimal bosonic Vasiliev models in $AdS_4$ have perturbative spectra given by a real scalar, a graviton and a tower of Fronsdal tensors of even ranks greater than two, which form an irreducible and unitarizable representation of the minimal higher spin Lie algebra extension $hs(4)$ of $so(3,2)\cong sp(4;\Real)$. While the linearization only refers to algebraic structures within the enveloping algebra of $sp(4;\Real)$, the full theory is based on the finer structure provided by the associative $\star$-product algebra generated by two Grassmann even $sp(4,\Real)$ quartets $(Y_{\ua},Z_{\ua})$, $\ua=1,...,4$, obeying 
\bea
\left[Y_{\ua},Y_{\ub}\right]_\star ~=~2iC_{\ua\ub}~=~-
\left[Z_{\ua}, Z_{\ub}\right]_\star\ ,\quad \left[Y_{\ua},Z_{\ub}\right]_\star ~=~0\ ,
\eea
and the reality conditions\footnote{An alternative convention is $(y^\a)^{\dagger}:=\bar y_{\ad}$ and $(\yb^{\ad})^\dagger:=-y_\a$, as used for example in an equivalent form in \cite{Chang:2012kt}.}
\be Y_{\ua}~:=~(y_{\a},\yb_{\ad})\ ,\quad Z_{\ua}~:=~(z_\a,-\zb_{\ad})\ ,\quad  (y_\a)^{\dagger}~:=~ \yb_{\ad}\ ,\quad (z_\a)^{\dagger}~:=~ \zb_{\ad}\ .\ee
Letting $x^\mu$ coordinatize a bosonic base manifold, which we refer to as $X$-space, and introducing anti-commuting line elements $(dx^\mu,dZ^{\ua})$ and  
\be 
\widehat d~ :=~  dx^\mu\del_\mu+ dZ^{\ua}\del_{\ua}\ ,\quad \del_{\ua}(\cdot)~:=~\frac{i}2 [Z_{\ua},\cdot]_\star\ ,
\ee
using conventions in which 
\be  
\widehat d(\widehat f\star\widehat g)~=~(\widehat  d \widehat f)\star\widehat g+(-1)^{{\rm deg}(\widehat f)}\widehat f\star(\widehat  d \widehat g)\ ,\quad (\widehat d \widehat f)^\dagger~=~\widehat d((\widehat f)^\dagger)\ ,\ee\be (\widehat f\star \widehat g)^\dagger~=~(-1)^{{\rm deg}(\widehat f)\,{\rm deg}(\widehat g)}(\widehat g)^\dagger\star (\widehat f)^\dagger\ ,
\label{formdagger}
\ee 
where ${\rm deg}$ denotes form degree, and writing $dz^2\equiv dz^\a\star dz_\a$ \emph{idem} $d\bar z^2$ and 
\be \left[\widehat f,\widehat g\right]_\pi~\equiv~ \widehat f\star\widehat g-(-1)^{{\rm deg}\widehat f{\rm deg}(\widehat g)}\widehat g\star\pi(\widehat f)\ ,\ee 
where the automorphism
\be 
\pi(\widehat f(y,\yb;z,\bar z))~:=~\widehat f(-y,\yb;-z,\bar z)\ ,\quad \pi \widehat d~=~ \widehat d\pi\ , 
\ee
\emph{idem} $\bar \pi$, the equations of motion for the minimal bosonic models read \cite{Vasiliev:1990en}\\
\bea
\widehat{\cal F}  &:=& \widehat d \widehat A +\widehat A \star \widehat A~=~ \frac{i}4 \left(dz^2 \widehat {\cal V}  + 
d\bar z^2 \widehat{\overline{\cal V}}\right)\ \ \ \ 
\label{master1}\w4
\widehat D\widehat\Phi  & :=&  \widehat d \widehat\Phi +\left[{\widehat A}, \widehat\Phi\right]_\pi = 0\ \ \ 
\label{master2}
\eea
where
\begin{itemize}
\item[(i)] the master one form 
$\widehat A  =  dx^\mu  {\widehat A}_{\mu}(x,Y,Z)+dZ^{\ua}\widehat A_{\ua}(x,Y,Z)$ and master zero form 
$\widehat \Phi=\widehat \Phi(x,Y,Z)$ are subject to the reality conditions 
\be
{\widehat A}^{\dagger}~=~ -\widehat A\ ,\quad {\widehat\Phi}^\dagger~=~\pi(\widehat\Phi)\ ,
\label{dagger}
\ee
and minimal bosonic projections
\be
\tau (\widehat A ) = - {\widehat A}\ ,\quad {\widehat A}^{\dagger}\se -\widehat A\ ,
\qquad
\t (\widehat\Phi )=  \pi (\widehat\Phi) \ ,\label{tau}
\ee
where the graded anti-automorphism 
\be
\t(\widehat f (Y,Z)) ~:=~ \widehat f(iY,-iZ)\ ,\quad \tau \widehat d~:=~ \widehat d\tau\ ,\ee
\be \tau(\widehat f\star \widehat g)~:=~(-1)^{{\rm deg}(\widehat f){\rm deg}(\widehat g)}\tau(\widehat g)\star \tau(\widehat f)\ ;\ee
\item[(ii)] the deformations
\be
\widehat {\cal V}~:=~\sum_{p=0}^\infty v_p (\widehat \Phi\star\widehat \kappa)^{\star p}\ ,\quad \widehat{\overline{\cal V}}~:=~(\widehat{\cal V})^\dagger\ ,\label{calV}\ee
where the Kleinians are defined by
\be 
\widehat\kappa\star \widehat f\star \widehat \kappa~:=~ \pi( \widehat f)\ ,\quad  \widehat{\bar\kappa}~:=~(\widehat\kappa)^\dagger\ ,
\ee
and $v_p=v_p(\widehat \Phi,\widehat A_{\ua})$ are $\tau$-invariant complex valued  functionals that are constant on shell, \emph{i.e.} $dv_p=0$ modulo the equations of motion; for non-trivial examples of such zero-form invariants, see \cite{Sezgin:2011hq,Colombo:2010fu,Boulanger:2011dd,Boulanger:2012bj}
\end{itemize}
Key features of the Vasiliev system are:
\begin{itemize}
\item its universal Cartan integrability\footnote{
 The universality plays a r\^ole in certain off-shell formulations \cite{Boulanger:2011dd,Boulanger:2012bj} and in HS geometries \cite{Sezgin:2011hq}.}, \emph{i.e.} its compatibility with $(\widehat d)^2\equiv 0$ on base manifolds of arbitrary dimension, which implies invariance under the gauge transformations
\be
\delta_{\widehat \e}\widehat A~=~\widehat D \widehat\epsilon~:=~  d+\left[\widehat A,\widehat \e\right]_\star\ ,\quad \widehat\delta_{\widehat\epsilon}\widehat\Phi ~=~-\left[\widehat\epsilon,\widehat\Phi\right]_\pi\ ,
\la{hfgt}
\ee
where $\widehat\e=\widehat\e(x,Y,Z)$ is subject to the same kinematic constraints as $\widehat A$. The closure relation $\left[\delta_{\widehat\e_1},\delta_{\widehat\e_2}\right]~=~\delta_{[\widehat\e_1,\widehat\e_2]_\star}$ defines the higher spin algebra $\widehat{hs}(4)$.  
\item its equivalent formulation as the deformed oscillator system
\be 
\widehat S_{[\a}\star\widehat S_{\b]}~=~-2i\e_{\a\b}(1-\widehat{\cal V})\ ,\quad \widehat{\overline S}_{[\ad}\star\widehat{\overline S}_{\bd]}~=~-2i\e_{\ad\bd}(1-\widehat{\overline{\cal V}})\ ,
\ee
\be 
\left[\widehat S_\a,\widehat S_{\ad}\right]_\star ~=~0\ ,\quad \left[ \widehat S_\a,\widehat\Phi\right]_{\bar\pi} ~=~0
\ ,\quad \left[ \widehat{\overline S}_{\ad},\widehat \Phi\right]_{\pi} ~=~0\ ,
\ee
where 
\be \widehat S_{\underline\a}~:=~ Z_{\underline\a}-2i\widehat A_{\underline\a}\ ,\label{Sa}\ee
coupled to the Maurer--Cartan system
\be \widehat F_{\mu\nu}~=~0\ ,\quad \widehat D_\mu \widehat \Phi~=~0\ ,\ee
\be \widehat D_\mu \widehat S_\a~:=~\partial_\mu \widehat S_\a+[\widehat A_\mu,\widehat S_\a]_\star~=~0\ .\ee
Splitting 
\be 
\widehat{\cal V}~=~\widehat{\cal V}_+ +\widehat{\cal V}_-\ ,\quad \widehat{\cal V}_\pm~=~\sum_{p=0}^\infty \frac12(1\pm (-1)^p) v_p (\widehat\Phi\star\widehat\kappa)^{\star p}\ ,
\ee
it follows from $\{\widehat S_\a,\widehat \Phi\star \widehat{\kappa}\}_\star=0=[\widehat {\overline S}_{\ad},\widehat\Phi\star\widehat\kappa]_\star$ that the perturbatively defined redefinition $\widehat S_{\a}\rightarrow \sqrt{1-\widehat{\cal V}_+}\star \widehat S_\a$ \emph{idem} $\widehat {\overline S}_{\ad}$ induces
\be 
\widehat {\cal V}_+~\rightarrow~0\ ,\quad \widehat{\cal V}_-~\rightarrow~ (1-\widehat{\cal V}_+)^{\star(-1)}\star\widehat{\cal V}_-\ .
\ee
By perturbatively redefining $\widehat\Phi$, one may take\footnote{The quantity $\widehat X$ obeys $D\widehat X := d\widehat X+\left[\widehat A,\widehat X\right]_\star=0$ and $(\widehat X)^\dagger=\tau(\widehat X)=\widehat X$. One also has $\widehat X=\widehat\Phi\star\bar\pi(\widehat\Phi)=\widehat\Phi\star\pi(\widehat\Phi)=(\widehat \Phi\star\widehat\kappa)^{\star 2}$.}
\be 
\widehat{\cal V}~=~\exp_\star(i\widehat\Theta)\star \widehat\Phi\star \widehat{\kappa}\ ,\quad \widehat\Theta~=~\sum_{p=0}^\infty \theta_p \widehat X^{\star p}\ ,\quad \widehat X~:=~(\widehat\Phi\star\widehat{\bar\k})^{\star2}\ ,\label{Theta}
\ee
where $\theta_p$ are $\tau$-invariant real valued zero-form invariants such that
\be 
\widehat \Theta~=~(\widehat \Theta)^\dagger~=~\tau(\widehat \Theta)\ ,\qquad 
\widehat D\widehat\Theta ~:= ~\widehat d\widehat\Theta+\left[\widehat A,\widehat \Theta\right]_\star~=~0\ ,
\label{Theta2}
\ee
modulo the equations of motion. 
\end{itemize}
The deformation $\widehat{\cal V}$ cannot be simplified further by perturbatively defined master field redefinitions \cite{Vasiliev:1992av}, which shows the key r\^ole played by the Kleinian operators\footnote{For example, replacing $\widehat{\cal V}(\widehat\Phi\star\widehat\kappa)$ by $\widehat{\cal V}(\widehat{\widetilde \Phi})$ where $\widehat{\widetilde \Phi}$ is an adjoint zero-form obeying $\widehat D\widehat{\widetilde\Phi}:=\widehat d \widehat{\widetilde\Phi}+[\widehat A,\widehat{\widetilde\Phi}]_\star=0$ yields a system without local perturbative degrees of freedom which can be brought to $\widehat F=0=\widehat D\widehat{\widetilde\Phi}$ by means of a perturbatively defined field redefinition}. As perturbatively defined redefinitions need not be globally defined in moduli space, the classification of non-perturbatively inequivalent Vasiliev systems remains an interesting open problem.

The minimal bosonic models are consistent truncations of non-minimal bosonic models obtained by replacing the $\tau$-projection by the weaker bosonic projection\footnote{We are assuming standard spin-statistics such that bosonic projection is tantamount to setting half-inter spins to zero.}
\be \tau^2(\widehat A)~\equiv~\pi\bar\pi(\widehat A)~=~\widehat A\ ,\qquad \tau^2(\widehat\Phi)~\equiv~\pi\bar\pi(\widehat\Phi)~=~\widehat \Phi\ ,\ee
and removing the $\tau$-invariance condition on $\theta_p$.

The parity map is the automorphism of the oscillator algebra defined by 
\be 
P(\widehat f(y,z;\bar y,\bar z))~=~\widehat f(\bar y,-\bar z;y,-z)\ ,\quad P d~=~ d P\ .\label{Pmap}
\ee
The $\widehat\Theta$-deformation breaks parity except in the following two cases \cite{Sezgin:2002ru}:
\bea 
\mbox{Bosonic Type A model}&:& \widehat\Theta~=~\theta_0~=~0\ ,\quad 
\\[5pt]
\mbox{Bosonic Type B model}&:& \widehat\Theta~=~\theta_0~=~\ft{\pi}2\ ,
\eea
for which the master field equations and kinematic constraints are invariant under
\be P(\widehat A)~=~\widehat A\ ,\quad P(\widehat \Phi)~=~e^{2i\theta_0}\widehat\Phi\ ,\ee
which assign intrinsic parity $e^{2i\theta_0}$ to the physical scalar. The parity invariant models may be minimal or non-minimal depending on whether the bosonic projection is imposed using $\tau$ or $\tau^2$.


\subsection{ De Sitter Space, Euclidean and Kleinian Signatures}


Bosonic models in four dimensional spacetimes with different signatures and different signs of the cosmological constant can be obtained by complexification of \eq{master1} and \eq{master2} by keeping the bosonic projection while dropping all reality conditions, and in particular treating $\widehat{\cal V}$ and  $\widehat{\overline{\cal V}}$ as independent odd $\star$-functions,  followed by imposition of suitably modified reality conditions;
for a detailed construction of these models, see \cite{Iazeolla:2007wt}, and for their harmonic expansions and spectra, see \cite{Iazeolla:2008ix}. 
The complexified HS algebra admits three distinct real forms, containing either $so(5)$, $so(4,1)$ and $so(3,2)$. 
Each of the latter are compatible with two different spacetime signatures, leading to five distinct models in total. The reality conditions read
\be \widehat A^\dagger~=~-\sigma(\widehat A)\ ,\qquad
\widehat\Phi^{\dagger}~=~\sigma(\pi(\widehat \Phi))\ ,\label{dagger}\ee
where the map $\sigma$ is given in Table \ref{Table1} and hermitian conjugates of doublets are defined as follows for different Lorentz algebras\footnote{In Euclidean signature, where we have absorbed the isomorphism $\rho$ used in \cite{Iazeolla:2007wt} into the symbol $\dagger$, one has $((\widehat f^\dagger))^\dagger\equiv\pi\bar\pi(\widehat f)$ and the reality conditions are consistent in view of the bosonic projection. 
}:
\bea su(2)_L\times su(2)_R&:&\quad (y^\a)^\dagger\ =\
y_{\a}\ ,\quad (z^\a)^\dagger\ =\ z_{\a}\
,\label{su2}\\&&\quad  (\bar y^{\ad})^\dagger\ =\ \bar
y_{\ad}\ ,\quad (\bar
z^{\ad})^\dagger\ =\ \bar z_{\ad}\ ,\nn\\[5pt]
su(2;\Comp)_{\rm diag}&:&\quad (y^\a)^\dagger\ =\ \bar y^{\ad}\
,\quad
(z^\a)^\dagger\ =\ \bar z^{\ad}\ ,\label{sl2}\\[5pt]
sp(2;\Real)_L\times sp(2;\Real)_R&:&\quad (y^\a)^\dagger\ =\ y^{\a}\
,\quad (z^\a)^\dagger\ =\ -z^\a\ ,\label{sp2}\\&&\quad (\bar
y^{\ad})^\dagger\ =\ \bar y^{\ad}\ ,\quad (\bar z^{\ad})^\dagger\ =\
-\bar z^{\ad} \ .\nn
\eea
The reality conditions in \eq{dagger} define the higher spin algebras $ho(p,5-p)\supset so(p,5-p)$, which only refer to the signatures of the isometry algebras, and their twisted adjoint representations, which also refer to the spacetime signatures. 
The signatures can be determined by using van de Waerden symbols to map $Y_{\underline{\a}} Y_{\underline{b}}$ into $M_{AB}$ obeying $(M_{AB})^\dagger=\sigma (M_{AB})$; for further details, see \cite{Iazeolla:2007wt,Iazeolla:2008ix}. 
As for the deformations, their reality conditions read 
\bea \mbox{Signature $(3,1)$}&:& \widehat{\cal V}^\dagger ~=~\widehat{\overline {\cal V}}\ ,\\
\mbox{Signatures $(4,0)$ and $(2,2)$}&:& \widehat{\cal V}^\dagger ~=~\widehat{\cal V}\ ,\qquad \widehat{\overline {\cal V}}^\dagger~=~ \widehat{\overline {\cal V}}\ ,\eea
and by re-defining $\widehat\Phi$, one may take

\be 
\mbox{Signature $(3,1)$}~:~\widehat {\cal V}~:=~e_\star^{i\widehat\Theta}\star \widehat \Phi\star\widehat\kappa\ ,
\qquad \widehat{\overline{\cal V}}~:=~({\cal V})^\dagger~=~e_\star^{-i\widehat\Theta}\star \widehat\Phi \star \widehat{\bar\kappa}\ ,
\ee
\be \mbox{Signature $(4,0)$ and $(2,2)$}~:~\widehat {\cal V}~:=~e_\star^{\widehat\Theta}\star \widehat \Phi\star\widehat\kappa\ ,\qquad \widehat{\overline{\cal V}}~:=~ \s_0 e_\star^{-\widehat{\Theta}}\star \widehat\Phi \star \widehat{\bar\kappa}\ ,
\ee

with $\widehat\Theta$ given by \eq{Theta} and $\s_0=\pm 1$. The parity assignments 
\bea 
P(\widehat A)&=&\widehat A\ ,\qquad P(\widehat\Phi)~=~ e^{2i\theta_0}
\widehat\Phi\ ,\quad \theta_0~=~0\,,\ \frac{\pi}2\ ,
\eea
with $P$ defined as in \eq{Pmap} for all signatures, imply that the physical scalar has intrinsic parity $e^{2i\theta_0}$ and that 
\bea 
\mbox{$(3,1)$ signature}&:& \widehat \Theta~=~\theta_0\ ,\\[5pt]
\mbox{$(4,0)$ and $(2,2)$ signature}&:& \widehat \Theta~=~0\ ,\quad \s_0~=~e^{2i\theta_0}\ .
\eea
In non-Lorentzian signatures, one has the maximally parity violating 
\be
\mbox{Chiral models ($(4,0)$ and $(2,2)$ signatures)
}~:~~~\widehat{\cal V}~=~\widehat\Phi\star\widehat\kappa\ ,\quad \widehat{\overline{\cal V}}~=~0\ .
\ee

{\footnotesize \tabcolsep=2mm  \begin{table}[h!]
\begin{center}
\begin{tabular}{|c|c|c|c|c|c|}
\hline
Isometry & Signature & Spinors & $\l^2$ &
$\sigma$ & Vacuum   \\ \hline  
\mbox{\footnotesize $so(5)$} & \mbox{\footnotesize $(4,0)$} &
\mbox{\footnotesize $SU(2)_L\times SU(2)_R$} & $-1$ & ${\rm id}$ &
\mbox{\footnotesize $S^4$} 
\\ \mbox{\footnotesize $so(4,1)$} & \mbox{\footnotesize $(4,0)$}
& \mbox{\footnotesize $SU(2)_L\times SU(2)_R$} & $+1$ & $\pi$
& \mbox{\footnotesize $H_4$}  \\
\mbox{\footnotesize $so(4,1)$} & \mbox{\footnotesize $(3,1)$} &
\mbox{\footnotesize $SL(2,\Comp)_{\rm diag}$} & $-1$ & $\pi$ &
\mbox{\footnotesize $dS_4$}   \\
\mbox{\footnotesize $so(3,2)$} & \mbox{\footnotesize $(3,1)$} &
\mbox{\footnotesize $SL(2,\Comp)_{\rm diag}$} & $+1$ & id &
\mbox{\footnotesize $AdS_4$}   \\
\mbox{\footnotesize $so(3,2)$} & \mbox{\footnotesize $(2,2)$} &
\mbox{\footnotesize $SL(2,\Real)_L\times SL(2,\Real)_R$} & $-1$ & id
& \mbox{\footnotesize $H_{2,2}$} \\  
\hline
\end{tabular}
\end{center}
\caption{\footnotesize Each row correspond to a real form of four dimensional higher spin gravity with indicated spacetime signature, cosmological constant $\Lambda=-3\lambda^2$ and vacuum solution with listed isometry. The spinor oscillators transform as doublets under the groups listed under spinors. $H_4$ stands for the 4-hyperboloid, also referred to as the Euclidean $AdS_4$, and $H_{2,2}$ stands for the coset $SO(3,2)/SO(2,2)$. }
\label{Table1}
\end{table}}


\section{Fermions and Yang--Mills Symmetries}\label{sec:exths}


Here we shall begin by reviewing the Konstein-Vasiliev construction of HS algebras that are extended by inclusion of algebras generated by matrices and fermionic elements, resulting in HS theories in $AdS_4$ with fermions and internal Yang--Mills symmetries. We shall then generalize these results to $dS_4$ as well as four dimensional spacetimes with Euclidean and Kleinian signatures.


\subsection{Konstein-Vasiliev Algebras in $AdS_4$}


Fermions and internal Yang--Mills symmetries can be introduced by tensoring the $(Y,Z)$-oscillator algebra by suitable matrix and Clifford algebras.
In the $AdS_4$ case, Konstein and Vasiliev \cite{Konstein:1989ij} have constructed three families of extended HS algebras admitting unitary representations given by squares of singletons and containing bosonic subalgebras given by the direct sums of $sp(4;\Real)\cong so(3,2)$ and the Yang-Mills algebras $u(m)\oplus u(n)$, $o(m)\oplus o(n)$ and $usp(m) \oplus usp(n)$, respectively, namely
\bea
hu(m;n|4))&:& S_\pm\otimes \bar S_\pm \ ,\label{singhu}\\[5pt]
ho(m;n|4)&:& \left[S_\pm\otimes S_\pm\right]_{\pm} \ ,\label{singho}\\[5pt]
husp(m;n|4)&:& \left[S_\pm\otimes S_\pm\right]_{\mp} \ ,\label{singhusp}
\eea
where $S_+:=(m,{\rm Rac})\oplus (n,{\rm Di})$ and $S_-:=(m,{\rm Di})\oplus (n,{\rm Rac})$, with Di and Rac referring to the spinor and scalar singleton representations of $sp(4;\Real)$, respectively, and $\left[\cdot\right]_{\pm}$ stand for symmetric and antisymmetric tensor products, respectively.
The singleton products decompose under the bosonic subalgebras as \cite{Konstein:1989ij}
\be
\begin{array}{lcll}
hu(m;n|4)&:& (m^2-1,1)\oplus (1,n^2-1)\oplus (1,1)\oplus (1,1)&s=0,1,2,3,\dots\\[5pt]
&& (m,\bar n)\oplus (\bar m,n)& s=\ft12,\ft32,\ft52,\dots\\[10pt]
ho(m;n|4)&:& (\ft12 m(m-1),1)\oplus (1,\ft12 n(n-1))&s=1,3,\dots\\[5pt]
&& (\ft12 m(m+1)-1,1)\oplus (1,\ft12 n(n+1)-1)\oplus (1,1)\oplus (1,1)&s=0,2,4,\dots\\[5pt]
&& (m, n)\oplus (m,n)& s=\ft12,\ft32,\ft52,\dots\\[10pt]
husp(m;n|4)&:& (\ft12 m(m+1),1)\oplus (1,\ft12 n(n+1))&s=1,3,\dots\\[5pt]
&& (\ft12 m(m-1)-1,1)\oplus (1,\ft12 n(n-1)-1)\oplus (1,1)\oplus (1,1)&s=0,2,4,\dots\\[5pt]
&& (m, n)\oplus (m,n)& s=\ft12,\ft32,\ft52,\dots\ ,
\end{array}\label{Konsteinspectra}
\ee
where the fields with spin $s\geqslant \ft12$ carry the representation $D(s+1;s)$ of the $AdS_4$ algebra with lowest energy $s+1$, as it should be for massless fields, and the scalars carry $D(1;0)$ or $D(2;0)$ depending on their intrinsic parity. The realization of these spectra in terms of linearized fields subject to suitable boundary conditions will be described below. The models with $mn>0$ contain fermions and are based on HS algebras that are superalgebras in the sense that they involve fermionic generators, but, as we shall review below, only a class of them contain the standard $AdS_4$ superalgebras.


\subsection{Full Models in (anti) de Sitter Space and Diverse Signatures}


Fully nonlinear interactions as well as the possibility of having different spacetime signatures and signs of the cosmological constant, are accounted for by extended master fields $(\widehat A,\widehat \Phi)$ valued in the direct product of the $(Y,Z)$-oscillator algebra and internal associative algebras given by 
\bea \mbox{$AdS_4$, $H_{2,2}$}&:& {\rm Cliff}_1(\Comp)\otimes {\rm Mat}_{m+n}(\Comp)\ ,\\[5pt]
\mbox{$dS_4$, $H_4$, $S^4$}&:& {\rm Cliff}_2(\Comp)\otimes {\rm Mat}_{m+n}(\Comp)\ ,\eea
where ${\rm Cliff}_k(\Comp)$ denotes the Clifford algebra with $k$ fermionic generators, say $\th^r$, $r=1,\dots,k$, obeying $\{\th^r,\th^s\}_\star=2\delta^{rs}$.
The master fields 
\begin{itemize}
\item[(i)] are Grassmann even, \emph{i.e.}
\be
\varepsilon_{\rm f}(\widehat A)~=~0~=~
\varepsilon_{\rm f}(\widehat \Phi)\ ,
\ee
where $\varepsilon_{\rm f}$ counts the Grassmann parity of generators as well as component fields; 
\item[(ii)] obey the spin-statistics conditions 
\be 
\pi\bar\pi \pi_\th(\widehat A)~=~ \widehat A\ ,\qquad \pi\bar\pi \pi_\th(\widehat \Phi)~=~\widehat \Phi\ ,\ee
where $\pi_\th$ is the automorphism with non-trivial action $\pi_\th(\th^r)=-\th^r$; and
\item[(iii)] belong to the graded matrix algebra defined by the projection 
\be \pi_\th {\rm Ad}_\Gamma(\widehat A)~=~ \widehat A\ ,\qquad \pi_\th {\rm Ad}_\Gamma(\widehat \Phi)~=~\widehat \Phi\ ,\qquad 
\Gamma~=~\left[\begin{array}{cc}\mathbf 1_{m\times m}&0\\0&-\mathbf 1_{n\times n}\ea\right]\ .
\ee
\end{itemize}
The further imposition of reality and $\tau$-conditions yields three families of models based on the adjoint and twisted adjoint representations of extended HS algebras as follows\footnote{The algebras $hu(m;n|2,3)$, $ho(m;n|2,3)$ and $husp(m;n|2,3)$, respectively, are isomorphic to $hu(m;n|4)$, $ho(m;n|4)$ and $husp(m;n|4)$.}:
\begin{itemize}
\item $hu(m;n|p,5-p)$: the master fields obey the reality conditions given in Table \ref{Table3} where the $\dagger$ acts in ${\rm Mat}_{m+n}(\Comp)$ as standard matrix hermitian conjugation and in the Clifford algebras as
\bea 
\mbox{$AdS_4$, $H_{2,2}$}&:& (\th^1)^\dagger~:=~\th^1\ ,\\[5pt]
\mbox{$dS_4$, $H_4$, $S^4$}&:& (\th^r)^\dagger~=~\e^{rs}\th^s\ ,
\eea
where the doublet structure is introduced so that the hermitian conjugation in the total $(Y,Z,\theta)$-oscillator algebra, which need not square to the identity, is compatible with the reality conditions;
\item $ho(m;n|p,5-p)$ and $husp(m;n|p,5-p)$: the master fields obey the further projections
\be 
\tau(\widehat A)~=~- \widehat A\ ,\quad \tau(\widehat \Phi)~=~ \bar\pi(\widehat \Phi)\ ,
\ee
where the graded anti-automorphism\footnote{By its definition, one has $\tau(\widehat f\star \widehat g) =(-1)^{\e(\widehat f)\e(\widehat g)+{\rm deg}(\widehat f){\rm deg}(\widehat g)} (\tau (\widehat f))\star (\tau(\widehat g))$.}
\be\tau(\widehat f(Y,Z,dZ,\theta))~:=~\left(\eta\star\widehat f(iY,-iZ,-idZ,i\theta)\star\eta^{-1}\right)^T\ ,\ee
with $(\cdot)^T$ denoting transposition in ${\rm Mat}_{m+n}(\Comp)$ and 
\bea
ho(m;n|p,5-p)&:& \eta~=~\left[\ba{cc} \mathbf 1_{m\times m}&0\\0& \mathbf 1_{n\times n}\ea\right]\ ,
\\[5pt]
husp(m;n|p,5-p)&:& \eta~=~\left[\ba{cc} \boldsymbol{\e}_{m\times m}&0\\0& \boldsymbol{\e}_{n\times n}\ea\right]\ ,
\eea
where $\boldsymbol{\e}_{m\times m}$ denotes the constant symplectic matrix of rank $m$. 
\end{itemize}
In all cases, the Vasiliev equations take the form \eq{master1} and \eq{master2} with 
\be \widehat{\cal V}~:=~\widehat{\cal V}\left(\widehat\Phi\star\widehat\kappa\Gamma\right)\ ,\quad
\widehat{\overline{\cal V}}~:=~\widehat{\overline{\cal V}}\left(\widehat\Phi\star\widehat{\bar\kappa}\right)\ ,\ee
which may be simplified using perturbatively defined field redefinitions, as discussed earlier, to obtain  
\bea \mbox{Signature $(3,1)$}&:&\widehat {\cal V}~=~e_\star^{i\widehat\Theta}\star \widehat \Phi\star\widehat\kappa\Gamma\ ,\qquad \widehat{\overline{\cal V}}~=~e_\star^{-i\widehat\Theta}\star \widehat\Phi \star \widehat{\bar\kappa}\ ,\label{calVGamma}\\[5pt] \mbox{Signature $(4,0)$ and $(2,2)$}&:&\widehat {\cal V}~=~e_\star^{\widehat\Theta}\star \widehat \Phi\star\widehat\kappa\Gamma\ ,\qquad \widehat{\overline{\cal V}}~=~\s_0 e_\star^{-\widehat{\Theta}}\star \widehat\Phi \star \widehat{\bar\kappa}\ ,\eea
with $\sigma_0=\pm 1$ and $\widehat\Theta$ given by \eq{Theta} where $\widehat X:=(\widehat \Phi\star\widehat{\bar\kappa})^{\star 2}=\widehat \Phi\star\bar\pi(\widehat\Phi)=\widehat \Phi\star\pi\pi_\th(\widehat\Phi)=(\widehat \Phi\star\widehat\kappa\Gamma)^{\star 2}$. 

In $AdS_4$ and $H_{2,2}$, the resulting set of dynamical Lorentz tensors and tensor-spinors is summarized in \eq{Konsteinspectra} with the understanding that $s$ refers to Lorentz spin.

The $dS_4$, $H_4$ and $S^4$ cases exhibit an additional doublet structure, \emph{viz.} 
\be \widehat A~\equiv~\sum_{\s=\pm} (\widehat A_\s + \widehat \Psi^i_\s \th^i )P_\s\ ,\quad 
\widehat \Phi~\equiv~\sum_{\s=\pm} (\widehat \Phi_\s + \widehat \chi^i_\s \th^i )P_\s\ ,\ee
where $P_\s:=\frac12(1\pm \Gamma_\th)$ with $\Gamma_\th:=i\th^1\star \th^2$ obeying $(\Gamma_\th)^\dagger=\Gamma_\th=-\tau(\Gamma_\th)$ and $(\Gamma_\th)^2=1$, implying that in models based on $hu(m;n|5-p,p)$ one has
\be (\widehat A_\s)^\dagger~=~ -\widehat A_\s\ ,\quad (\widehat \Psi^i_\s)^\dagger~=~\e^{ij}\widehat \Psi^j_{-\s}\ ,\ee
while in models based on $ho(m;n|5-p,p)$ and $husp(m;n|5-p,p)$ one has the additional $\tau$ conditions
\be \tau(\widehat A_\s)~=~-\widehat A_{-\s}\ ,\quad \tau(\widehat \Psi^i_\s)~=~-i\widehat \Psi^i_{\s}\ .
\ee

Assigning parities as  
\bea 
P(\widehat A)&=&\widehat A\ ,\qquad P(\widehat\Phi)~=~ e^{2i\theta_0}
\widehat\Phi\star \Gamma\ ,\label{Passign}
\eea
with $P$ defined as in \eq{Pmap} for all signatures,
parity invariance requires that 
\be \widehat\Theta~=~ \theta_0~=~0\,,\ \frac{\pi}2\ ,\quad \sigma_0~=~e^{2i\theta_0}\ ,\label{AandB}\ee
which we refer to as the Type A and Type B models with fermions and Yang--Mills symmetries, respectively. 
The dynamical scalars in $\widehat\Phi|_{Y=Z=0}$ can be arranged into real scalars with definite intrinsic parities as follows:
\bea \mbox{$AdS_4$, $dS_4$}&:& \widehat\Phi|_{Y=Z=0}~:=~\phi_++i\phi_-+(\phi_+-i\phi_-)\star\Gamma\ ,\\[5pt]
 \mbox{$H_{2,2}$, $S^4$, $H_4$}&:& \widehat\Phi|_{Y=Z=0}~:=~\phi_+\star(1+\Gamma)+\phi_-\star(1-\Gamma)\ ,\eea
where
\be (\phi_\pm)^\dagger~=~\phi_\pm\ ,\quad P(\phi_\pm)~=~\pm e^{2i\theta_0} \phi_\pm\ .\ee
%


\subsection{Linearization and Spectrum}


In all cases, letting $(\Phi,W)$ be the fluctuations in $(\widehat \Phi,\widehat A)|_{Z=0}$ around a spin $2$ background $\langle (\widehat \Phi,\widehat A)\rangle = (0,\Omega)$ obeying
\be  d\O+\O\star\O~=~0\ ,\quad \O~:=~e+\omega\ ,\label{BGeq}\ee
\be e~:=~\frac1{2i} e^{\a\ad}y_\a \yb_{\ad}\ ,\quad \omega~:=~\frac1{4i}\left(\omega^{\a\b} y_\a y_\b +\bar\o^{\ad\bd}\yb_{\ad}\yb_{\bd}\right)\ ,\ee
where $e^{\a\ad}=\frac{\l}2 (\s_a)^{\a\ad} e^a$ with $\l^2>0$ for $AdS_4$, $H_{2,2}$ and $S^4$, and $\l^2<0$ for $dS_4$ and $H_4$, and further details are given in Appendix \ref{App:2},
the linearized equations of motion read
\be \nabla W+\{e,W\}_\star+\frac{i}2 \left (e^{-i\theta_0}e^{\a\ad} \bar e_{\ad}{}^\b \frac{\partial^2}{\partial y^\a\partial y^\b} \Phi|_{\yb=0}+e^{i\theta_0}\bar e^{\ad\a} e_{\a}{}^{\bd}\frac{\partial^2}{\partial \yb^{\ad}\partial\yb^{\bd}} \Phi|_{y=0}\star \Gamma\right)~=~0\ ,\label{lin1}\ee
\be \nabla\Phi+\{e,\Phi\}_\star~=~0\ ,\label{lin2}\ee
where $\nabla=d+{\rm ad}_{\omega}$ denotes the background Lorentz covariant derivative.
As Eqs. \eq{BGeq}, \eq{lin1} and \eq{lin2} have been obtained by linearizing a nonlinear Cartan integrable system, it follows on general grounds that they are left invariant under linearized Cartan gauge symmetries, \emph{viz.}
\be 
\delta_\e e~=~\nabla\e_\x+[e,\e_\L]_\star\ ,\quad \delta_\e \omega~=~\nabla\e_\L+[e,\e_\xi]_\star\ ,\ee
\be \delta_\e W~=~\nabla \e+[e,\e]_\star+[W,\e_\Lambda]_\star+[W,\e_\xi]_\star-i \left (e^{-i\theta_0}\e_\x^{\a\ad} \bar e_{\ad}{}^\b \frac{\partial^2}{\partial y^\a\partial y^\b} \Phi|_{\yb=0}+{\rm h.c.}\right)\ ,\label{Weq}\ee
\be \delta_\e \Phi~=~[\Phi,\e_\Lambda]_\star-\{\Phi,\e_\xi\}_\star\ ,
\ee
as well as nonabelian Killing symmetries with parameters $\e_0$ valued in the Konstein--Vasiliev algebra at $Z^{\ua}=0$ obeying 
\be 
\nabla  \e_0+[e,\e_0]_\star~=~0\ ,
\ee
and acting on the fluctuation fields in accordance with the structure of quadratic terms  in nonlinear system (see Appendix B)
which  in the case of the Weyl zero-form amounts to 
\be \delta_{\e_0}\Phi~=~-[\e_0,\Phi]_\pi\ ,\ee
while $\delta_\l W=[W,\l]_\star+{\rm trilinear\ terms}$; see Appendix B. 
By means of algebraic Cartan integration, \emph{viz.} $(\O,W,\Phi)=(\exp \overrightarrow T') (\O',W',\Phi')|_{W'=0,\,\O'=0}$ where $\overrightarrow T'$ is the generator of Cartan gauge transformations with finite gauge functions and $\Phi'$ is a constant twisted-adjoint element, one has
\be \Omega~=~L^{-1}\star dL\ ,\quad  \Phi~=~L^{-1}\star \Phi'\star \bar\pi(L)\ ,\ee
where $L$ is a background gauge function, and $W$ given algebraically in terms of $\Phi'$, $L$ and its own gauge functions.
Likewise, the Killing parameters 
\be 
\e_0~=~L^{-1}\star \e'_0\star L\ .\ee
Hence, taking $\e_0'$ and $\Phi'$ to belong to adjoint and twisted-adjoint representations of  the KV algebra, respectively, and choosing a gauge function $L$, one obtains expansions of the dynamical fields and Killing parameters in terms of harmonic functions on the maximally symmetric background, which by construction obey the lineraized field equation subject to boundary conditions that are consistent with their forming representations of the nonabelian KV algebra.
In particular, in $AdS_4$ and $dS_4$, this method can be applied using unitary twisted-adjoint representations for massless fields; for a treatment of $dS_4$, see \cite{Iazeolla:2008ix}, and for mixed-symmetry fields in diverse dimensions, see \cite{Boulanger:2008up,Boulanger:2008kw}.

In $AdS_4$, one may consider harmonic expansions in which the fields with $s\geqslant \ft12$ carry lowest weight spaces $D(s+1;s)$ of $so(3,2)$ with intrinsic parity $(-1)^s$ for bosons; the compact basis element with energy $\o$ and spin $m$ corresponds to a harmonic Weyl zero-form $\Phi^{s+1;s;\o,m}_{a(s),b(s)}(x^\mu)$ that is an eigenfunction of the $AdS_4$ Killing vectors corresponding to the energy operator and a compact spin generator. 
Since the spin $s$ Weyl tensor of the spin $s$ gauge field is given by 
\be C_{\a(s)}~=~e^{-i\theta_0} \Phi_{\a(2s)}\ ,\ee
it follows that if one introduces the electro-magnetic characteristic
\footnote{More generally, the generalized Petrov classes for $\Phi_{a(s),b(s)}$ are labelled by ordered partitions $\{n_1,\dots,n_k\}$ of $2s$ characterized as $\Phi_{\a(2s)}=\nu_s \prod_{i=1}^k (u^i_\a)^{n_i}$ where $u^i_\a$ are $k$ non-collinear polarization spinors.}
\be
(-1)^{s+1}\Phi_{a(s),b(s)}\Phi^{a(s),b(s)}~\left\{\ba{ll}~ >~0& \mbox{magnetic}\ ,\\[5pt]
~<~0& \mbox{electric}\ ,\ea\right.
\ee
\emph{idem} $C_{a(s),b(s)}$, then the characteristics of $C_{a(s),b(s)}$ and $\Phi_{a(s),b(s)}$ are the same for sufficiently small $\theta_0$ and opposite for $\theta_0$ sufficiently close to $\ft{\pi}2$ \cite{Iazeolla:2011cb}.
As for the scalars, unbroken higher spin symmetry requires $\phi_\pm$ to carry the lowest weight spaces $D(2,0)_-$ and $D(1,0)_+$ for $\theta_0=0$ and vice versa for $\theta_0=\ft\pi 2$.

Instead of using lowest-weight spaces, the linearized fields can be expanded in $\theta_0$-dependent bases induced by Fronsdal tensors with point-like magnetic sources at the Minkowskian boundary of the Poincar\'e coordinate chart in $AdS_4$ \cite{Chang:2012kt}. 
The resulting linear solution space in the twisted-adjoint module, corresponding to unfolded bulk-to-boundary propagators with the aforementioned magnetic boundary conditions, is spanned by Weyl zero-forms $\Phi^{\theta_0;\vec x,\chi}(x^\m)$ labelled by points $\vec x$ in three-dimensional Minkowski spacetime and real polarization spinors $\chi_\a$.
As found in \cite{Chang:2012kt}, if $0<\theta_0<\ft{\pi}2$, this solution space, in general, is not left invariant by all Killing transformations, though supersymmetric HS models were constructed in which the solution space is preserved by finite dimensional $AdS_4$ superalgebras for ${\cal N}=1,2,3,4,6$. We shall comment on these models further below.

{\footnotesize \tabcolsep=2mm  \begin{table}[h!]
\begin{center}
\begin{tabular}{|c|c|c|c|c|c|}
\hline
&&&&&\\
Vacuum & $(\widehat A)^\dagger$ & $(\widehat \Phi)^\dagger$ & $((\widehat A)^\dagger)^\dagger$ & $((\widehat \Phi)^\dagger)^\dagger$ & $((\cdot)^\dagger)^\dagger$  \\[5pt] 
\hline 
&&&&&\\
$AdS_4$ & $-\widehat A$ & $\pi(\widehat \Phi)\star\Gamma$ & $\widehat A$ & $\pi\bar\pi\pi_\th(\widehat \Phi)$ & ${\rm id}$ \\[7pt]
$H_{2,2}$ & $-\widehat A$ & $\bar\pi(\widehat \Phi)$ & $\widehat A$ & $\widehat \Phi$ & ${\rm id}$ \\[7pt]
$dS_4$ & $-\pi(\widehat A)$ & $\widehat \Phi\star\Gamma$ & $\pi\bar\pi(\widehat A)$ & $\pi_\th(\widehat \Phi)$ & $ \pi_\th$ \\[7pt]
$H_4$ & $-\bar\pi(\widehat A)$ & $\widehat \Phi$ & $\widehat A$ & $\widehat \Phi$ & $\pi\bar\pi\pi_\th$ \\[7pt]
$S^4$ & $-\widehat A$ & $\bar\pi(\widehat \Phi)$ & $\widehat A$ & $\widehat \Phi$ & $ \pi\bar\pi\pi_\th$ \\[5pt]
\hline
\end{tabular}
\end{center}
\caption{\footnotesize The reality conditions on the adjoint and twisted adjoint master fields in different signatures, as indicated by the corresponding vacua, are tabulated in the second and third columns. The nesting of the reality conditions yields the automorphisms listed in the fourth and fifth columns, while $\dagger^2$ is identically equal to the automorphisms given in the last column. Thus, in all signatures except that of $H_{2,2}$, the consistency of the reality condition requires the spin-statistics projection $\pi\bar\pi\pi_\th(\widehat A,\widehat \Phi)=(\widehat A,\widehat \Phi)$. The reality conditions are also consistent with the identity $\tau((\widehat f)^\dagger)\equiv (\tau(\widehat f))^\dagger$.  }
\label{Table3}
\end{table}}


\section{Supersymmetric Higher Spin Theories }


In this Section we describe a subset of the KV models which are based on HS extension of ordinary AdS superalgebras. These will fall into three families with ${\cal N}=1,2,4$ mod $4$, respectively, while the minimal models with ${\cal N}=3$ mod $4$ automatically have ${\cal N}=4$ mod $4$ as we shall show below.
Next, for a given value ${\cal N}$, we extend the HS theories by tensoring the underlying HS algebras with internal symmetries. Finally, we spell out some details of the minimal ${\cal N}=2$ model in $dS_4$, which has the minimum amount of supersymmetry allowed in $dS_4$.


\subsection{Even ${\cal N}$ in $AdS_4$}


In the $AdS_4$ case, the extended models based on Konstein--Vasiliev algebras are supersymmetric in the usual sense iff $m=n$, i.e. if the underlying supersingleton contains equal number of bosons and fermions, in which case
\bea 
ho(m;m|4)&\supset& osp(1|4)~\ni~ Q_{\a} ~:=~ \th^1 y_\a \otimes \sigma^1\otimes \mathbf{1}_{m\times m} \ ,\\[5pt]
hu(m;m|4)&\supset& osp(2|4)~\ni~ Q_{\a}^{\pm} ~:=~ \th^1 y_\a \otimes (\sigma^1\pm i\s^2)\otimes \mathbf{1}_{m\times m} \ ,\\[5pt]
husp(m;m|4)&\supset& osp(4|4)~\ni~ Q^I_{\a} ~:=~ \th^1 y_\a \otimes \Sigma^I\otimes \mathbf{1}_{\frac{m}2\times \frac{m}2} \ ,
 \eea
where, in the last case, $\eta={\bf 1}_{2\times 2}\otimes i\sigma^2\otimes \mathbf{1}_{\frac{m}2\times \frac{m}2}$ and $\Sigma^I=(\sigma^2 \otimes \s^i,\sigma^1\otimes \bf 1_{2\times 2})$, $I=1,\dots,4$, are $so(4)$ gamma matrices. 

If $m=n=2^{k}$ then 
\be \frac12 \left(1+\pi_\th {\rm Ad}_\Gamma\right)\left( {\rm Cliff}_1(\Comp)\otimes {\rm Mat}_{m+n}(\Comp) \right) ~\cong~ {\rm Cliff}_{{\cal N}}(\Comp)\ ,\quad {\cal N}~=~2(k+1)\ ,\ee
with generators $\xi^i$, $i=1,\dots,{\cal N}$, obeying 
\be \{\xi^i,\xi^j\}_\star~=~2\delta^{ij}\ ,\quad (\xi^i)^\dagger~=~\xi^i\ ,\quad \tau(\xi^i)~=~\left\{\ba{ll} i\xi^i&\mbox{${\cal N}=4$ mod $4$}\ ,\\[5pt]
-i\xi^1\star \xi^i\star \xi^1&\mbox{${\cal N}=2$ mod $4$}\ ,\ea\right.\ee
and one can identify 
\be \Gamma~\cong~\Gamma_\xi~:=~i^{k+1}\xi^1\star\cdots\star\xi^{\cal N}\ .\ee
Thus, the sequence of models based on the minimal higher spin extensions $shs^E({\cal N}|4)$ of $osp({\cal N}|4)$ in which the master fields $(\widehat A,\widehat \Phi)$ are valued in the $(Y,Z,\xi)$-oscillator algebra and obey  (see Table \ref{Table:SusyModels} )
\begin{itemize}

\item[(i)] spin-statistics, \emph{viz.}
\be \varepsilon_{\rm f}(\widehat A,\widehat \Phi)~=~(0,0)\ ,\quad \pi\bar\pi\pi_\xi(\widehat A,\widehat\Phi)~=~(\widehat A,\widehat \Phi)\ ,\ee
where $\pi(\xi^i):=-\xi^i$;

\item[(ii)] the reality conditions
\be (\widehat A)^\dagger~=~-\widehat A\ ,\quad (\widehat\Phi)^\dagger~=~\pi(\widehat\Phi)\star \Gamma\ ;\label{real}\ee

\item[(iii)] additional $\tau$-conditions for 

\be \mbox{${\cal N}=4$ mod $4$}~:~~\tau(\widehat A)~=~-\widehat A\ ,\quad \tau(\widehat \Phi)~=~\bar\pi(\widehat \Phi)\star\Gamma \ ;\label{tau}\ee
\item[(iv)] the Vasiliev equations \eq{master1} and \eq{master2} with deformations as in \eq{calVGamma} including the parity invariant $shs^E({\cal N}|4)$ models of Type A and Type B defined by \eq{Passign} and \eq{AandB};
\end{itemize}
is equivalent to sequences of models based on isomorphic Konstein--Vasiliev algebras as follows \cite{Konstein:1989ij} (${\cal N}=2(k+1)$):
\be 
shs^E({\cal N}|4)~{\cong}~\left\{\ba{ll} hu(2^k;2^k|4)& k=0,2,\dots\ ,\\[5pt]
husp(2^{k}; 2^{k}|4)& k=1,5,\dots
\\[5pt] 
ho(2^{k}; 2^{k}|4)& k=3,7,\dots\ .\ea\right.
\ee
The chain of consistent truncations from ${\cal N}$ to ${\cal N}-2$ can be made manifest by reformulating the subsequence of $shs^E({\cal N}|4)$ models with ${\cal N}=2$ mod $4$ by introducing an additional ${\rm Cliff}_2(\Comp)$ algebra with fermionic generators $\eta^r$, $r=1,2$, obeying
\be 
\{\eta^r,\eta^s\}~:=~\delta^{rs}\ ,\quad \varepsilon_{\rm f}(\eta^r)~=~0\ ,\quad (\eta^r)^\dagger~:=~\eta^r\ ,\quad \tau(\eta^r)~:=~i\eta^r\ ,
\ee
and re-defining  (see Table \ref{Table:SusyModels} )
\be 
\mbox{${\cal N}=2$ mod $4$}~:~~ \tau(\xi)~:=~i\xi^i\ ,\quad \Gamma~:=~\Gamma_\xi\star\Gamma_\eta\ ,\quad 
\Gamma_\eta~:=~i\eta^1\star \eta^2\ .
\ee
The minimal ${\cal N}=2$ mod $4$ models can then be formulated equivalently using master fields $(\widehat A,\widehat \Phi)$ valued in the $(Y,Z,\xi,\eta)$-oscillator algebra subject to spin-statistics
\be 
\varepsilon_{\rm f}(\widehat A,\widehat\Phi)~=~(0,0)\ ,\quad \pi\bar\pi\pi_\xi\pi_\eta(\widehat A,\widehat\Phi)~=~(\widehat A,\widehat \Phi)\ ,
\ee
where $\pi_\eta(\eta^r)=-\eta^r$; the reality conditions \eq{real}; the $\tau$-conditions \eq{tau}; and the additional $Z_2$-projection
\be 
[\Gamma_\eta,\widehat A]_\star~=~0~=~[\Gamma_\eta,\widehat \Phi]_\star\ .
\ee
The Vasiliev equations take the form \eq{master1} and \eq{master2} with deformations as in \eq{calVGamma}, and the parity invariant minimal ${\cal N}=2$ mod $4$ models of Type A and Type B are defined by \eq{Passign} and \eq{AandB}.

The $shs^E(8|4)$ model was analyzed in detail in \cite{Sezgin:1998gg} and its truncation to minimal models with ${\cal N}=1,2,4$ was described in \cite{Engquist:2002vr}\footnote{Models with ${\cal N}=1,2,4$ have also appeared recently in \cite{Chang:2012kt}, which, however, are not minimal as they are all formulated using four Clifford algebra generators and not imposing any $\tau$ condition.}   The minimal ${\cal N}=6$ model, given recently in \cite{Chang:2012kt}, can be obtained by first splitting $\hat\xi^{\hat i}=(\xi^{i},\eta^r)$ where $\hat i=1,\dots,8$, $i=1,\dots,6$ and $r=1,2$ and then imposing
\be  
\left[\Gamma_\eta,\widehat A\right]_\star~=~0\ ,\qquad  \left[\Gamma_\eta,\widehat \Phi\right]_\star~=~0\ ,
\ee
which eliminates the gravitino supermultiplet and its higher spin analogs, resulting in the spectrum of the minimal ${\cal N}=6$ model given in Table \ref{Table2}. As shown above, this model can equivalently be formulated \cite{Chang:2012kt} using only the fermionic $\xi^i$ oscillators, provided one drops the $\tau$ condition; compare the second and third rows of Table \ref{Table:SusyModels}.

\begin{table}[t]
\begin{center}
{\footnotesize \tabcolsep=1mm
\begin{tabular}{|c|cccccccccccccc|}\hline
& & & & & & & & & & & & &
& \\
{\large${}_{\ell}\backslash s$} & $0$ & \ns{$\ft12$} & $1$ &
\ns{$\ft32$} & $2$ & \ns{$\ft52$} & $3$ & \ns{$\ft72$} & $4$ &
$\ft92$ & $5$ & $\ft{11}2$ &
$6$ & $\cdots$ \\
& & & & & & & & & & & & & & \\ \hline & & & & & & & & & & & & &
& \\
$0$ & $15+\overline{15}$ & $20+6$ & $15+1$ & $6$ & $1$ &
& & & & & & & & \\
$1$ & $1+\bar 1$ & $6$ & $15+1$ & $20+6$ & $15+15$ & $20+6$ & $15+1$ & $6$ &
$1$ &
& & & &  \\
$2$ & & & & & $1$ & $6$ & $15+1$ & $20+6$ & $15+15$
& $20+6$ & $15+1$ & $6$ & $1$ &  \\
$3$ & & & & & & & & & $1$ & $6$ & $15+1$ & $20+6$ & $15+15$
& $\cdots$ \\
$4$ & & & & & & & & & & & &
& $1$ & $\cdots$ \\
$\vdots$ & & & & & & & & & & & & & &  \\ \hline
\end{tabular}}
\end{center}
\caption{\footnotesize The table contains the $SO(3,2)\times SO(6)$ content 
of the minimal ${\cal N}=6$ model in $4D$ arranged into $osp(6|4)$ 
supermultiplets labelled by $\ell$ with the entries referring to 
$SO(6)$ irreps with the supergravity multiplet at $\ell=0$ .
The spin $1$ sector of the supergravity multiplets arises by gauging $R$-symmetry 
generators $T^{ij}=\xi^i\xi^j$ and the central generator $\Gamma_\eta$.
} \la{Table2}
\end{table}


\subsection{Odd ${\cal N}$ in $AdS_4$}


The sequence of minimal ${\cal N}=1$ mod $4$ models, including the ${\cal N}=1$ model of \cite{Engquist:2002vr}, based on the extended HS algebras $shs^E({\cal N}|4)$ with  maximal finite-dimensional subalgebras $osp({\cal N}|4)$, has master fields $(\widehat A,\widehat \Phi)$ depending on $(Y^{\ua},Z^{\ua}\xi^i,\eta)$, where $(\xi^i,\eta)$, $i=1,\dots,{\cal N}$ are fermionic generators of ${\rm Cliff}_{{\cal N}+1}(\Comp)$,  obeying spin-statistics, \emph{i.e.}
\be \varepsilon_{\rm f}(\widehat A,\widehat\Phi)~=~(0,0)\ ,\quad \pi\bar\pi\pi_\xi\pi_\eta(\widehat A,\widehat\Phi)~=~(\widehat A,\widehat \Phi)\ ;\label{ssxieta}\ee
the reality conditions \eq{real} where
\be \mbox{${\cal N}=1$ mod $4$}~:~~\Gamma~:=~ i \xi^1\star\cdots\star\xi^{\cal N}\star\eta\ ;\ee
the $\tau$-conditions \eq{tau} using
\be \tau(\xi^i)~:=~i\xi^i\ ,\quad \tau(\eta)~:=~-i\eta\ ,\ee
which implies $\tau(\Gamma)=\Gamma$. The equations of motion are of the standard format \eq{master1} and \eq{master2} with deformations as in \eq{calVGamma} and parity invariant minimal models of Type A and Type B defined as in \eq{Passign} and \eq{AandB}.

In particular, the minimal $\cN=1$ model is based on the HS algebra $shs(1|4)$ whose maximal finite-dimensional subalgebra is $osp(1|4)$. The spectrum of the $shs(1|4)$ gauge theory is given by the symmetric product of two $osp(1|4)$ singletons, which decomposes into a tower of $\cN=1$ massless supermultiplets with $s_{max}= \ell +\frac12$ and  $s_{\rm max}= 2\ell+2$ for $\ell=0,1,2...$, thus containing, in particular, a Wess-Zumino scalar multiplet and the supergravity multiplet. 

In the case of ${\cal N}=3$ mod $4$, the analog of the above construction requires fermionic oscillators $(\xi^i,\eta^r)$, $i=1,\dots,{\cal 
N}$, $r=1,2,3$, obeying $\tau(\xi^i,\eta^r)=(i\xi^i,-i\eta^r)$ in order for there to exist the element
\be \mbox{${\cal N}=3$ mod $4$}~:~~ \Gamma~:=~i\xi^1\star\cdots\star\xi^{\cal N}\star
\eta^1\star\eta^2\star\eta^3\ ,\quad \Gamma\star\Gamma~=~1\ ,\quad (\Gamma)^
\dagger~=~\tau(\Gamma)~=~\Gamma\ .\ee
The resulting minimal model, in which the master fields obeys \eq{ssxieta}, \eq{real} and \eq{tau} together with the further projection
\be \mbox{${\cal N}=3$ mod $4$}~:~~ [T^{rs},\widehat A]_\star~=~=~[T^{rs},\widehat
\Phi]_\star\ ,\quad T^{rs}~:=~\eta^r\eta^s\ , \ee
which is consistent as $[T^{rs},\Gamma]_\star=0$, thus has ${\cal N}+1$ supersymmetries generated 
by $Q^i_{\ua}=\xi^i Y_{\ua}$ and $Q^{{\cal N}+1}_{\ua}=i\eta^1 \eta^2\eta^3 Y_{\ua}$. Thus, ${\cal N}=3$ mod $4$ implies 
${\cal N}=4$ mod $4$ in HS theories.


\begin{table}[t]
\begin{center}
{\footnotesize \tabcolsep=1mm
\begin{tabular}{|c|c|c|c|c|}\hline
&&&&\\
${\cal N}$ mod $4$  &\  Fermionic\  \ & $\Gamma$ & \ Projections \ \ & Internal   \\
 &oscillators && & symmetry\\ 
 &&&&algebra\\[5pt]
\hline 
&&&&\\
$1$  & $\xi^i,\eta$ & $i\xi^1\cdots \xi^{\cal N}\eta$ & $\tau$ &\  $o(n)$ or $usp(n)$\  \ \\[5pt]
$2$  & $\xi^i,\eta^r$ & \ $\xi^1\cdots \xi^{\cal N}\eta^1\eta^2$ \ \ & $\tau$, $\Gamma_\eta$ & $u(n)$  \\[5pt]
$2$  & $\xi^i$ &  $i\xi^1\cdots \xi^{\cal N}$ & --- & $u(n)$  \\[5pt]
$4$  & $\xi^i$ &  $\xi^1\cdots \xi^{\cal N}$ & $\tau$  & $o(n)$ or $usp(n)$\\[5pt]
\hline
\end{tabular}}
\end{center}
\caption{\footnotesize The table displays the basic features of the three sequences of supersymmetric models in $AdS_4$ in accordance with ${\cal N}$ standard supersymmetries: The second and thirds columns list the fermionic oscillators and corresponding $\Gamma$ operators, respectively. In all models, the reality conditions read $(\widehat A,\widehat \Phi)^\dagger=(-\widehat A,\pi(\widehat\Phi)\star \Gamma)$. 
The $\tau$-conditions, when imposed as indicated in the fourth column, read $\tau(\widehat A,\widehat \Phi)=\eta\star(-\widehat A,\bar\pi(\widehat\Phi))\star\eta^{-1}$ where $\tau(\xi^i)=i\xi^i$ and $\tau(\eta^r)=-i\eta^r$, and $\eta=\mathbf 1_{n\times n}$ for $o(n)$ and $\eta=\mathbf \e_{n\times n}$ for $usp(n)$. 
Extending a given minimal model without increasing ${\cal N}$ yields the internal symmetry algebras listed in the last column. The ${\cal N}=2$ mod $4$ models admit two equivalent formulations; the one with additional fermionic oscillators requires a $\tau$ condition and a further projection $[\Gamma_\eta,\widehat A]_\star=0=[\Gamma_\eta,\widehat \Phi]_\star$ where $\Gamma_\eta:= i\eta^1 \eta^2$. 
} \label{Table:SusyModels}
\end{table}


\subsection{Internal Symmetries in $AdS_4$}


The minimal models in $AdS_4$ with ${\cal N}=1,2,4$ mod $4$ supersymmetries can be extended by internal symmetries without changing ${\cal N}$ by first removing the reality and $\tau$ conditions on the master fields and then tensoring them with an internal ${\rm Mat}_n(\Comp)$ algebra, after which the reality and $\tau$ conditions can be reimposed with $\dagger $ and $\tau$ acting on matrices as in Sections 4.1 and 4.2. 
In the case of ${\cal N}=1,2,4$, the spectra can be read off from the following expansions of the now matrix-valued Weyl zero-form:
\bea 
{\cal N}~=~1&:& \Phi|_{\yb=0}~=~M(y)+\Gamma M'(y)+\xi \Psi(y)+\eta \Psi'(y)\ ,
\label{expansions1}\\[5pt]
{\cal N}~=~2&:& \Phi|_{\yb=0}~=~M(y)+\Gamma M'(y) + \xi^i \Psi_i(y)\ ,\\[5pt]
{\cal N}~=~4 &:&  \Phi|_{\yb=0}~=~ M(y)+\xi^i\star\xi^jM_{[ij]}(y)+\Gamma M'(y)+\xi^i \Psi_i(y)+ \xi^i\star\xi^j\star\xi^j \Psi_{[ijk]}(y)\  , \qquad\quad
\label{expansions3}
\eea
where $\Phi={\widehat \Phi}|_{Z=0}$. The resulting spectra are given in Table \ref{TableExtendedSpectra}. 
In the case of ${\cal N}=2$, for which there are two types of realizations as described in Section 4.1, the corresponding spectra are the same. 
Furthermore, the reality properties of the fields shown in Table \ref{TableExtendedSpectra} can be determined from \eq{real}. For example, in the scalar sector, $M^\dagger|_{y=0}= M'|_{y=0}$ for ${\cal N}=1,2,4$, and in the case of ${\cal N}=4$, we have $M_{ij}^\dagger|_{y=0}=\frac12\epsilon_{ijkl} M^{k\ell}|_{y=0}$. 

The spectrum analysis proceeds similarly for all ${\cal N}=1,2,4$ mod $4$ as well. In particular, in the case of  ${\cal N}=2$ mod $4$, for which there are two types of realizations as described in Section 4.1, the corresponding spectra are the same. For example, in the case of ${\cal N}=6$, the spectrum is the one given in Table 3, with every field is taken to be $u(n)\sim su(n)\oplus u(1)$ valued. The $u(1)$ part contains the ${\cal N}=6$ supergravity multiplet at the lowest level.

\begin{table}[t]
\begin{center}
{\footnotesize \tabcolsep=1mm
\begin{tabular}{|c|c|c|c|c|c|}\hline
&&&&&\\
\ Supersymmetry\  & \ Internal\  & $s=0$ mod $2$ & $s=1$ mod $2$ & $s=\ft12$ mod $2$ & $s=\ft32$ mod $2$ \\ 
&&&&&\\
\hline
&&&&&\\
${\cal N}=1$ & $o(n)$ & $\SY\oplus \SY'$ & $\AS\oplus \AS'$ & $\AS\oplus \SY'$ & $\SY\oplus \AS'$ \\[5pt]
 & $usp(n)$ & $\AS\oplus \AS'$ & $\SY\oplus \SY'$ & $\SY\oplus \AS'$ & $\AS\oplus \SY'$ \\
 &&&&&\\
 \hline
 &&&&&\\
 ${\cal N}=2$  & $u(n)$ & $\MAT\oplus\MAT'$ & $\MAT\oplus\MAT'$ & $\MAT{}^i$ & $\MAT{}^i$ \\
 &&&&&\\
 \hline
 &&&&&\\
 ${\cal N}=4$  & $o(n)$ & $\SY\oplus\AS{}^{ij}\oplus \SY'$ & $\AS\oplus\SY{}^{ij}\oplus \AS'$ & $\AS{}^i\oplus \SY{}^{ijk}$ & $\SY{}^i\oplus \AS{}^{ijk}$ \\[5pt]
  & $usp(n)$ & $\AS\oplus\SY{}^{ij}\oplus \AS'$ & $\SY\oplus\AS{}^{ij}\oplus \SY'$ & $\SY{}^i\oplus \AS{}^{ijk}$ & $\AS{}^i\oplus \SY{}^{ijk}$ \\
  &&&&&\\
 \hline

\end{tabular}}
\end{center}
\caption{\footnotesize The spectra of ${\cal N}=1,2,4$ supersymmetric HS theories with internal symmetry. In the case of $u(n)$ and $usp(n)$ internal symmetries, the Young tableaux refer to the symmetry properties of the matrix valued quantities in \eqref{expansions1} and \eqref{expansions3}. In the case of $u(n)$ internal symmetry, the Yang tableaux with crosses denote ${\rm Mat}_n (\Comp)$ matrices. The maximal finite dimensional supersubalgebra is given by the direct sum of $osp({\cal N}|4)$ and the internal symmetry algebra.}
\label{TableExtendedSpectra}
\end{table}


\subsection{ ${\cal N}=2$ in $dS_4$ }


In $dS_4$, the ${\cal N}=2$ supersymmetric HS theory is based on the algebra $ho(1;1|4,1)$ contained in the general construction given above. As in de Sitter supergravity, ${\cal N}=2$ is the smallest possible number of supersymmetries.
In what follows, we shall elaborate further on this case and provide an alternative description in terms of oscillators alone.
The master fields belong to the associative algebra given by ${\rm Mat}_{1+1}(\mathbf{C})$ times the oscillator algebra generated by $(y^\alpha,\bar y^{\dot\alpha},z^\alpha,\bar z^{\dot\alpha},\theta^r)$ where the fermionic doublet obeys
\be
\{\theta^r,\theta^s\}_\star~=~2\delta^{rs}\ ,\quad (\theta^r)^\dagger~=~\epsilon^{rs}\theta^s\ .
\ee
The hermitian conjugation is defined as usual on the matrices and on the twistor oscillators in accordance with $SL(2,\mathbf{C})$ invariance. As a result
\be
((\widehat f)^\dagger)^\dagger~\equiv~ \pi_\theta(\widehat f)\ .
\ee
This is consistent with imposing spin-statistics projection 
\be
\epsilon_{\rm f}(\widehat A)~=0~=~\epsilon_{\rm f}(\widehat \Phi)\ ,\quad \pi\bar\pi\pi_\theta(\widehat A,\widehat \Phi)~=~(\widehat A,\widehat \Phi)\ ,\quad \Gamma\star (\widehat A,\widehat \Phi)\star \Gamma~=~(\widehat A,\widehat \Phi)\ , 
\ee
where $\epsilon_{\rm f}$ denotes the Grassmann parity and $\Gamma~:=~\sigma^3$.  The reality conditions leading to a de Sitter vacuum are
\be
\widehat A^\dagger~=~-\pi(\widehat A)\ ,\quad \widehat \Phi^\dagger~=~\widehat \Phi\star \Gamma\ .
\ee
In the $ho(1;1|4,1)$ model the master fields also obey the $\tau$ condition
\be \tau(\widehat A)~=~-\widehat A\ ,\quad \tau(\widehat\Phi)~=~\bar\pi(\widehat\Phi)\ ,\ee
where $\tau$ acts on $M_{1+1}(\Comp)$ by transposition. 
The supercharges of the ${\cal N}=2$ $dS_4$ supersymmetry algebra are realized as
\be
Q_\alpha^{r}~:=~y_\alpha \theta^r \sigma^1\ ,\quad \bar Q_{\dot\alpha}^{r}~:=~
\bar y_{\dot\alpha} \theta^r\sigma^1\ ,
\ee
obeying $(Q_\a^r)^\dagger=\e^{rs}Q^s_{\ad}$ and 
\be
\{Q^{r}_\alpha,Q^{s}_{\beta}\}_\star~=~2\delta^{rs} M_{\alpha\beta}+2i\epsilon_{\alpha\beta} T^{rs}\ ,
\quad\quad  \{Q^{r}_\alpha,Q^{s}_{\dot\beta}\}_\star~=~2\delta^{rs} P_{\alpha\dot\beta}\ ,
\ee
where $T^{rs}:= \frac12 [\theta^r,\theta^s]_\star$ is the $so(2)_R$ generator. 

The above model can equivalently be realized in terms of master fields depending on $(Y^{\ua},\xi^i)$, $i=1,2$ and where $\xi^i$ are fermionic Clifford algebra generators obeying 
\be (\xi^i)^\dagger~=~ \e^{ij}\xi^j\ .\ee
The reality conditions read
\be (\widehat A^\dagger,\widehat\Phi)^\dagger~=~(-\widehat A,\widehat \Phi\star\Gamma)\ ,
 \Gamma ~=~ i\xi^1\star \xi^2\ ,\ee
while there are no $\tau$ conditions. The supersymmetry charges read
\be Q^i_\a~=~ \xi^i y_\a\ ,\quad \overline Q_{\ad}^i~=~\xi^i \yb_{\ad}\ .\ee

The set of dynamical fields coincides with that of the ${\cal N}=2$ model in $AdS_4$ 
though the reality conditions on the fermions are modified. It is well known that ${\cal N}=2$ supergravity in $4D$ contains a vector ghost \cite{Pilch:1984aw}. It would be interesting to determine the corresponding situation in the HS version of the theory we have presented here.


\section{Comments}


The results on supersymmetric HS theories described here, old and new, are hoped to play a role in understanding their relation to string/M theory. 
Moreover, HS theories in  $dS_4$, Euclidean and Kleinian spacetimes provide  fertile grounds for sharpening ideas in holography, in the case of de Sitter space providing a framework in which problems that are notoriously difficult to study in the usual string theory can now be addressed \cite{Anninos:2011ui,Ng:2012xp}. Investigations on HS holography, attempts to make connection with string/M theory and the need to understand better the already existing interaction deformation in parity non-invariant HS theories are likely to motivate further generalizations of HS theories. 
It would be interesting, for example, to construct matter couplings systematically. Three dimensional Chern-Simons-quiver theories, which are holographically dual to the Freund-Rubin compactifications of M theory to $AdS_4$ (see, for example, \cite{Martelli:2008rt,Martelli:2008si}), in appropriate limits may be relevant for such constructions. In this context, the ${\cal N}=3$ compactification of M theory on $AdS_4\times N^{010}$ has been considered  briefly  in \cite{Engquist:2002vr}. In the case of ${\cal N}=1$ HS theories, the problem of constructing chiral matter couplings would obviously be of great interest. 
 

In analyzing certain aspects of a wide class of supersymmetric HS theories covered here, it may be useful to formulate them in superspace. Such a formulation is conceptually simple and mathematically manageable, given the universal Cartan integrable nature of Vasiliev equations. Indeed, starting with the standard formulation of Vasiliev equations in $AdS_4$, they can be formulated in superspace simply by replacing the 4D spacetime with a $D=(4|4\cN)$ superspace with $4\cN$ anti-commuting $\theta$-coordinates \cite{Engquist:2002gy}. This introduces extra spinorial directions in the $1$-forms as well as $\theta$-dependence in all component fields. On the other hand, there are also new constraints coming from projections of the differential form constraints in the new spinorial directions. As a result, each supermultiplet in the spectrum is described by a single constrained superfield, and we arrive at a superspace description of HS theory in AdS$_4$ which is equivalent to the formulation in ordinary spacetime. This procedure has been performed in \cite{Engquist:2002gy} for ${\cal N}=1$ supersymmetric HS theory in $AdS_4$. 


In their current form, Vasiliev equations for HS theories in four dimensions contain an infinite set of free parameters of which a finite number show up at every new order in perturbation theory.  
All of these parameters break parity when they are non-vanishing, except the first parameter, denoted by $\theta_0$, which preserves parity when it takes the values $0$ or $\pi/2$, for which the physical scalar has intrinsic parity $+1$ and $-1$, respectively.
The parameter $\theta_0$ shows up already at the free level, 
where it can be redefined away, however, by a duality rotation 
that mixes electric and magnetic components of the Weyl tensors.
At cubic order, $\theta_0$ remains the only free parameter, and it has been proposed in \cite{Giombi:2011kc} and tested in \cite{Giombi:2011kc,Aharony:2011jz} that this deformation parameter corresponds to the 't Hooft coupling of  Chern--Simons theories in $3D$ coupled to singletons in such a way that free scalars go into free fermions at strong coupling and vice versa; for a recent review, see \cite{Giombi:2012ms}.

So far, the nature of the three-dimensional counterparts of the higher-order deformation parameters remains less clear.
In \cite{Boulanger:2011dd,Boulanger:2012bj} it was found that the off-shell formulation based on generalized Hamiltonian actions requires the deformation function to be linear, hence containing only the $\theta_0$ parameter, albeit under certain extra assumptions that remain to be validated.
On the other hand, as pointed out in \cite{Sezgin:2011hq}, at the classical level, the set of deformations can be enriched even further by replacing each deformation parameter by a zero-form charge \cite{Sezgin:2005pv,Iazeolla:2007wt}.
The properties of the perturbative expansions of zero-form charges found in \cite{Colombo:2010fu,Colombo:2012jx} suggest that these new deformations could correspond to extensions of the generating function in three dimensions by 
additions of composite operators coupled to nonlinear sources.
In particular, beyond ${\cal N}=8$ such extensions would resolve some issues related to the absence of an ${\cal N}=8$ barrier for 4D HS supergravities with singlet gravitons and generally covariant weak field expansions with $AdS_4$ vacua,
though the presence of the $\theta_0$ parameter for ${\cal N}\geqslant 8$ remains a puzzle in the context of holography since Chern-Simons-matter type CFTs do not exist in three dimensions with such supersymmetries. 

The singletons play a key r\^ole in HS gravity in $AdS_4$, as they can be realized as unitary representations of the HS algebra using either the $Y^{\ua}$ oscillators or free fields in three dimensions.
These two dual realizations carry over to other signatures with the key difference that they are no longer unitary.
In particular, in $dS_4$ the massless bulk fields can be expanded harmonically in unitarizable representations consisting of states that are tensors of the maximal compact $so(4)\subset so(4,1)$.
These representations are not of lowest-weight type, but rather generalized Harish-Chandra modules: the analog of the ground state is thus the smallest $so(4)$ irrep, from which the remainder of the representation space can be generated by acting with the coset $so(4,1)/so(4)$.
In fact, the $so(4)$ content of the unitarizable representation for a massless spin $s$ field in $dS_4$ is exactly the same as the $so(3,1)$ content of its twisted-adjoint representation \cite{Iazeolla:2008ix}.
It would thus be interesting to characterize group theoretically the free scalar and fermion on $S^3$, thought of as a boundary of $dS_4$, and examine how the square of these representations, which are nonunitary, can be rearranged into the unitary irreps for the massless fields, that is, to generalize to $dS_4$ the Flato--Fronsdal formula.

In higher dimensions, in order for supersymmetric models to admit standard spacetime interpretations in $AdS_D$, one must have $D\leqslant 7$.  Higher spin models with $32$ supercharges in $D=5,7$ have been constructed at the linearized level, including their twisted adjoint representations in 
\cite{Sezgin:2001yf,Sezgin:2002rt}. These constructions rely on Grassmann even spinor oscillators $Y^{\ua}$ and Grassmann odd Clifford algebra generators. An interesting open problem is the construction of corresponding fully nonlinear models.

Breaking of HS symmetries is another clearly important problem. A mechanism for breaking of HS symmetries \cite{Girardello:2002pp}, which is well understood at the kinematic level in the case of bosonic models, involves Goldstone bosons as composite of a scalar and spin $(s-2)$ field for giving mass to a spin $s$ field. Some aspects of this mechanism has been discussed in \cite{Leigh:2003gk} for the ${\cal N}=1$ HS theory, and it would be useful to explore this further. Given that there is no ${\cal N}=8$ barrier in writing down fully nonlinear and consistent HS theories, it would also be interesting to determine whether in such models the symmetries can be broken down to ${\cal N}\leqslant 8$ supersymmetric HS gravity or ordinary gravity by a GPZ-like mechanism \cite{Girardello:2002pp} or by the mechanism proposed in \cite{Chang:2012kt}.


\section*{Acknowledgements}

We acknowledge each others home institutions and Bosphorus University for hospitality extended to us during visits. We also thank N. Boulanger and N. Colombo for useful discussions. The research of E.S. is supported in part by NSF grant PHY-0906222.


\begin{appendix}


\section{Spinor Conventions in Different Signatures}\label{App:2}


We use spinor conventions in which doublet indices are raised and lowered as follows:
\be y^\a~=~\epsilon^{\a\b}y_\b\ ,\quad y_\a~=~y^\b\epsilon_{\b\a}\ ,\quad \e^{\a\b}\e_{\c\d} \ = \ 2 \d^{\a\b}_{\c\d} \ , \qquad
\e^{\a\b}\e_{\a\c} \ = \ \d^\b_\c \ .\ee
The van der Waerden
symbols $(\s^a)_{\a\bd}\equiv (\sb^a)_{\bd\a}$ obey
 \bea
  (\s^{a})_{\a}{}^{\ad}(\sb^{b})_{\ad}{}^{\b}&=& \y^{ab}\d_{\a}^{\b}\
 +\ (\s^{ab})_{\a}{}^{\b} \ ,\qquad
 (\sb^{a})_{\ad}{}^{\a}(\s^{b})_{\a}{}^{\bd}\ =\ \y^{ab}\d^{\bd}_{\ad}\
 +\ (\sb^{ab})_{\ad}{}^{\bd} \ ,\label{so4a}\w2
 \ft12 \e_{abcd}(\s^{cd})_{\a\b}&=& \e (\s_{ab})_{\a\b}\ ,\qquad \ft12
 \e_{abcd}(\sb^{cd})_{\ad\bd}\ =\  -\e (\sb_{ab})_{\ad\bd}\ ,\label{so4b}
\eea
where $\e=\sqrt{\det \eta_{ab}}$. 
The reality conditions on doublet variables are summarized in
(\ref{su2}), (\ref{sl2}) and (\ref{sp2}) and for the van der Waerden symbols we use
\be (\e_{\a\b},(\s^a)_{\a\bd},(\s^{ab}_{\a\b}))^\dagger \ = \ \left\{ \begin{array}{ll}
(\e^{\a\b},-(\s^a)^{\a\bd},(\s^{ab})^{\a\b})& \mbox{for $SU(2)$}\ , \\[5pt]
(\e_{\ad\bd},(\s^a)_{\b\ad},(\bar\s^{ab})_{\ad\bd})& \mbox{for $SL(2,\Comp)$}\ ,  \\[5pt]
(\e_{\a\b},(\s^a)_{\a\bd},(\s^{ab})_{\a\b}) & \mbox{for $Sp(2)$}\ , \end{array} \right. \ee
corresponding to the following representations:
\bea  SU(2)&: &  \qquad \s^a=(i,\s^i) \ ,\qquad \sb^a=(-i,\s^i)  \
,\qquad
\e=i\s^2 \ ; \\[5pt]
 SL(2,\Comp)&: & \qquad \s^a=(-i\s^2,-i\s^i\s^2) \ ,\qquad
\sb^a=(-i\s^2,i\s^2\s^i) \ , \qquad\e=i\s^2 \ ; \\[5pt]
Sp(2)& : & \qquad \s^a=(1,\tilde{\sigma}^i) \ , \qquad\sb^a=(-1,\tilde{\sigma}^i) \
,\qquad \e=i\s^2 \ ,\eea
where $\tilde{\sigma}^i := (\s^1,i\s^2,\s^3)$. The real forms of $so(5;\Comp)$ are defined by \cite{Iazeolla:2007wt} 
\be [M_{AB},M_{CD}]\ =\ i\y_{BC}M_{AD}+\mbox{$3$ more}\ ,\qquad
(M_{AB})^\dagger\ =\ \sigma(M_{AB})\ ,\label{sogena}\ee
where $\eta_{AB}=(\eta_{ab};-\lambda^2)$. The commutation relations
decompose into
\be [M_{ab},M_{cd}]_\star\ =\ 4i\y_{[c|[b}M_{a]|d]}\ ,\qquad
[M_{ab},P_c]_\star\ =\ 2i\y_{c[b}P_{a]}\ ,\qquad [P_a,P_b]_\star\ =\
i\lambda^2 M_{ab}\ .\label{sogenb}\ee
The corresponding oscillator realization is taken to be
 \be
 M_{ab}\ =\ -\frac18 \left[~ (\s_{ab})^{\a\b}y_\a y_\b+
 (\sb_{ab})^{\ad\bd}\bar y_{\ad}\yb_{\bd}~\right]\ ,\qquad P_{a}\ =\
 \frac{\l}4 (\s_a)^{\a\bd}y_\a \yb_{\bd}\ .\label{mab}
 \ee
The real form of the
$so(5;\Comp)$-valued connection $\O$ we choose as
 \be
  \O\ =\ \frac1{4i}
 dx^\mu\left[\omega_\mu^{\a\b}~y_\a y_\b
 +\bar{\omega}_\mu{}^{\dot\a\dot\b}~{\bar y}_{\dot\a}{\bar y}_{\dot\b}
 + 2 e_\mu^{\a\dot\b}~y_\a {\bar y}_{\dot\b}\right]\
 ,\label{Omega}
 \ee
where
 \be
 \o^{\a\b}\ =\ -\ft14(\s_{ab})^{\a\b}~\o^{ab}\ ,
 \quad
 \bar{\omega}^{\dot\a\dot\b}\ =\ -\ft14({\sb}_{ab})^{\dot\a\dot\b}~\o^{ab}\ ,
 \quad
 e^{\a\dot\a}\ =\ \ft{\lambda}2(\s_{a})^{\a \dot\a}~e^{a}\ .
 \label{convert}
 \ee
Likewise, for the curvature ${\cal R}=d\O+\O\wedge\star \O$ one
finds
 \bea
 {\cal R}_{\a\b}&=& d\o_{\a\b} +\o_{\a\c}\wedge\o_{\b}{}^{\c}+
 e_{\a\dd}\wedge e_{\b}{}^{\dd}\ ,
 \label{rab}\w2
 \bar{\cal R}_{\dot\a\dot\b}&=& d\bar{\o}_{\ad\bd}
 +\bar{\o}_{\ad\cd}\wedge\bar{\o}_{\bd}{}^{\cd}
 +e_{\d\ad}\wedge e^{\d}{}_{\bd}\ ,
 \label{radbd}\w2
 {\cal R}_{\a\dot\b}&=&  de_{\a\bd}+ \o_{\a\c}\wedge
 e^{\c}{}_{\bd}+\bar{\o}_{\bd\dd}\wedge e_{\a}{}^{\dd}\
 ,\label{rabd}
 \eea
and
 \be
 {\cal R}^{ab}\ =\ d\o^{ab}+\o^a{}_c\wedge\o^{cb} +\lambda^2
 e^a\wedge e^b\ ,\qquad {\cal R}^a\ =\ d e^a+\o^a{}_b\wedge e^b\ .
 \label{curvcomp} \ee
%


\section{Linearized Symmetries}


The generalized curvature constraint 
\be
dX^\a+Q^\a(X)=0\ ,
\ee
where $Q^\a$ obey the Cartan integrability condition $Q^\a\partial_\a Q^\b=0$,
is invariant under gauge transformations 
\be
\delta_\e X^\a=T^\a(X,\e):=d\e^\a-\e^\b \partial_\b Q^\a\ .
\ee
If $\overline X^\a$ is a background then so is 
$X^\a(\overline X,v):= (\exp \overrightarrow T)\overline X^\a$, where $\overrightarrow T:=T^\b(\bar X,v)\bar\partial_\b$ and $v^\a$ are finite gauge 
functions; \emph{c.f.} normal coordinates. The locally defined moduli space thus consists 
of the gauge orbits over the constant solutions, \emph{i.e.} of elements $X^\a(\overline X,\l)$ with $\overline X^\a=\delta_{p_\a,0} C^\a$ where $p_\a:={\rm deg}(X^\a)$ and $dC^\a=0$. Expanding 
\be
X^\a=\overline X^\a +x^\a\ ,
\ee
yields 
\bea
&& dx^\a+x^\b \bar\partial_\b  \overline Q^\a+\frac12 x^\b x^\c \bar\partial_
\c \bar\partial_\b \overline Q^\a+\cdots=0\ ,
\nn\w2
&& \delta_\e x^\a=d\e_1^\a-\e_1^\b\bar\partial_\b\overline Q^\a-\e_0^\b x^\c \bar\partial_\c\bar\partial_\b\overline Q^\a+\cdots\ .
\eea
Thus, the linearized limit $dx^\a+x^\b \bar\partial_\b  \overline Q^\a=0$ is Cartan integrable with abelian local symmetries 
\be
\delta_{\e_1} x^\a=d\e_1^\a-\e_1^\b\bar\partial_\b \overline Q^\a\ ,
\ee
as well as nonabelian global symmetries
\be
\delta_{\e_0}x^\a=-\e_0^\b x^\c \bar\partial_\c \bar\partial_\b \overline Q^\a\ ,
\ee
for parameters obeying the Cartan--Killing equation $d\e_0^\a-\e_0^\b\bar\partial_\b \overline Q^\a=0$.

\end{appendix}

\providecommand{\href}[2]{#2}\begingroup\raggedright\endgroup


\begin{thebibliography}{10}

\bibitem{Vasiliev:1990en}
M.~A. Vasiliev, ``{Consistent Equation For Interacting Gauge Fields of All
  Spins In (3+1)-Dimensions},'' {\em Phys.Lett.} {\bf B243} (1990)
378--382.

\bibitem{Vasiliev:1995dn}
M.~A. Vasiliev, ``{Higher Spin Gauge Theories in Four Dimensions, Three
  Dimensions, and Two Dimensions},'' {\em Int.J.Mod.Phys.} {\bf D5} (1996)
  763--797,
\href{http://arXiv.org/abs/hep-th/9611024}{{\tt hep-th/9611024}}.

\bibitem{Vasiliev:1999ba}
M.~A. Vasiliev, ``{Higher Spin Gauge Theories: Star Product and AdS Space},''
  \href{http://arXiv.org/abs/hep-th/9910096}{{\tt hep-th/9910096}}.
Contributed article to Golfand's Memorial Volume, M. Shifman ed., World
  Scientific.

\bibitem{Bekaert:2005vh}
X.~Bekaert, S.~Cnockaert, C.~Iazeolla, and M.~Vasiliev, ``{Nonlinear Higher
  Spin Theories in Various Dimensions},''
\href{http://arXiv.org/abs/hep-th/0503128}{{\tt hep-th/0503128}}.

\bibitem{Sezgin:2002rt}
E.~Sezgin and P.~Sundell, ``{Massless Higher Spins and Holography},'' {\em
  Nucl.Phys.} {\bf B644} (2002) 303--370,
\href{http://arXiv.org/abs/hep-th/0205131}{{\tt hep-th/0205131}}.

\bibitem{Klebanov:2002ja}
I.~Klebanov and A.~Polyakov, ``{AdS Dual of the Critical O(N) Vector Model},''
  {\em Phys.Lett.} {\bf B550} (2002) 213--219,
\href{http://arXiv.org/abs/hep-th/0210114}{{\tt hep-th/0210114}}.

\bibitem{Bergshoeff:1988jm}
E.~Bergshoeff, A.~Salam, E.~Sezgin, and Y.~Tanii, ``{Singletons, Higher Spin
  Massless States and the Supermembrane},'' {\em Phys.Lett.} {\bf B205} (1988)
237.

\bibitem{Bergshoeff:1988jx}
E.~Bergshoeff, A.~Salam, E.~Sezgin, and Y.~Tanii, ``{N=8 Supersingleton Quantum
  Field Theory},'' {\em Nucl.Phys.} {\bf B305} (1988)
497.

\bibitem{Duff:1987qa}
M.~Duff, ``{Supermembranes: The First Fifteen Weeks},'' {\em Class.Quant.Grav.}
  {\bf 5} (1988)
189.

\bibitem{Bergshoeff:1987dh}
E.~Bergshoeff, M.~Duff, C.~Pope, and E.~Sezgin, ``{Supersymmetric Supermembrane
  Vacua And Singletons},'' {\em Phys.Lett.} {\bf B199} (1987)
69.

\bibitem{Blencowe:1987bn}
M.~Blencowe and M.~Duff, ``{Supersingletons},'' {\em Phys.Lett.} {\bf B203}
  (1988)
229.

\bibitem{Nicolai:1988ek}
H.~Nicolai, E.~Sezgin, and Y.~Tanii, ``{Conformally Invariant Supersymmetric
  Field Theories on $S^p\times S^1$ and Super p-Branes},'' {\em Nucl.Phys.}
  {\bf B305} (1988)
483.

\bibitem{Flato:1978qz}
M.~Flato and C.~Fronsdal, ``{One Massless Particle Equals Two Dirac Singletons:
  Elementary Particles in a Curved Space. 6.},'' {\em Lett.Math.Phys.} {\bf 2}
  (1978)
421--426.

\bibitem{Sezgin:1998gg}
E.~Sezgin and P.~Sundell, ``{Higher Spin N=8 Supergravity},'' {\em JHEP} {\bf
  9811} (1998) 016,
\href{http://arXiv.org/abs/hep-th/9805125}{{\tt hep-th/9805125}}.

\bibitem{Sezgin:1998eh}
E.~Sezgin and P.~Sundell, ``{Higher Spin N=8 Supergravity in AdS(4)},''
\href{http://arXiv.org/abs/hep-th/9903020}{{\tt hep-th/9903020}}.

\bibitem{Leigh:2003gk}
R.~G. Leigh and A.~C. Petkou, ``{Holography of the N=1 Higher Spin Theory on
  $AdS_4$},'' {\em JHEP} {\bf 0306} (2003) 011,
\href{http://arXiv.org/abs/hep-th/0304217}{{\tt hep-th/0304217}}.

\bibitem{Sezgin:2003pt}
E.~Sezgin and P.~Sundell, ``{Holography in 4D (Super) Higher Spin Theories and
  a Test Via Cubic Scalar Couplings},'' {\em JHEP} {\bf 0507} (2005) 044,
\href{http://arXiv.org/abs/hep-th/0305040}{{\tt hep-th/0305040}}.

\bibitem{Chang:2012kt}
C.-M. Chang, S.~Minwalla, T.~Sharma, and X.~Yin, ``{ABJ Triality: From Higher
  Spin Fields to Strings},''
\href{http://arXiv.org/abs/1207.4485}{{\tt 1207.4485}}.

\bibitem{Fradkin:1986ka}
E.~Fradkin and M.~A. Vasiliev, ``{Candidate to the Role of Higher Spin
  Symmetry},'' {\em Annals Phys.} {\bf 177} (1987)
63.

\bibitem{Vasiliev:1986qx}
M.~A. Vasiliev, ``{Extended Higher Spin Superalgebras and Their Realizations in
  Terms of Quantum Operators},'' {\em Fortsch.Phys.} {\bf 36} (1988)
33--62.

\bibitem{Fradkin:1987ah}
E.~Fradkin and M.~A. Vasiliev, ``{Superalgebra of Higher Spins And Auxiliary
  Fields},'' {\em Int.J.Mod.Phys.} {\bf A3} (1988)
2983.

\bibitem{Konstein:1989ij}
S.~Konstein and M.~A. Vasiliev, ``{Extended Higher Spin Superalgebras and Their
  Massless Representations},'' {\em Nucl.Phys.} {\bf B331} (1990)
475--499.

\bibitem{Konshtein:1988yg}
S.~Konshtein and M.~A. Vasiliev, ``{Massless Representations and Admissibility
  Condition for Higher Spin Superalgebras},'' {\em Nucl.Phys.} {\bf B312}
  (1989)
402.

\bibitem{Engquist:2002vr}
J.~Engquist, E.~Sezgin, and P.~Sundell, ``{On N=1, N=2, N=4 Higher Spin Gauge
  Theories in Four Dimensions},'' {\em Class.Quant.Grav.} {\bf 19} (2002)
  6175--6196,
\href{http://arXiv.org/abs/hep-th/0207101}{{\tt hep-th/0207101}}.

\bibitem{Sezgin:2001yf}
E.~Sezgin and P.~Sundell, ``{Towards Massless Higher Spin Extension of D=5, N=8
  Gauged Supergravity},'' {\em JHEP} {\bf 0109} (2001) 025,
\href{http://arXiv.org/abs/hep-th/0107186}{{\tt hep-th/0107186}}.

\bibitem{Vasiliev:2004cm}
M.~Vasiliev, ``{Higher Spin Superalgebras in Any Dimension and Their
  Representations},'' {\em JHEP} {\bf 0412} (2004) 046,
\href{http://arXiv.org/abs/hep-th/0404124}{{\tt hep-th/0404124}}.

\bibitem{Vasiliev:2004cp}
M.~Vasiliev, ``{Higher Spin Gauge Theories in Any Dimension},'' {\em Comptes
  Rendus Physique} {\bf 5} (2004) 1101--1109,
\href{http://arXiv.org/abs/hep-th/0409260}{{\tt hep-th/0409260}}.

\bibitem{Vasiliev:1992ix}
M.~A. Vasiliev, ``{Equations of Motion for D = 3 Massless Fields Interacting
  Through Chern-Simons Higher Spin Gauge Fields},'' {\em Mod.Phys.Lett.} {\bf
  A7} (1992)
3689--3702.

\bibitem{Vasiliev:1992av}
M.~A. Vasiliev, ``{More on Equations of Motion for Interacting Massless Fields
  of All Spins in (3+1)-Dimensions},'' {\em Phys.Lett.} {\bf B285} (1992)
225--234.

\bibitem{Sezgin:2002ru}
E.~Sezgin and P.~Sundell, ``{Analysis of Higher Spin Field Equations In Four
  Dimensions},'' {\em JHEP} {\bf 0207} (2002) 055,
\href{http://arXiv.org/abs/hep-th/0205132}{{\tt hep-th/0205132}}.

\bibitem{Sezgin:2011hq}
E.~Sezgin and P.~Sundell, ``{Geometry and Observables in Vasiliev's Higher Spin
  Gravity},''
\href{http://arXiv.org/abs/1103.2360}{{\tt 1103.2360}}.

\bibitem{Colombo:2010fu}
N.~Colombo and P.~Sundell, ``{Twistor Space Observables and Quasi-Amplitudes in
  4D Higher Spin Gravity},'' {\em JHEP} {\bf 1111} (2011) 042,
\href{http://arXiv.org/abs/1012.0813}{{\tt 1012.0813}}.

\bibitem{Boulanger:2011dd}
N.~Boulanger and P.~Sundell, ``{An Action Principle for Vasiliev's
  Four-Dimensional Higher Spin Gravity},'' {\em J.Phys.A} {\bf A44} (2011)
  495402, \href{http://arXiv.org/abs/1102.2219}{{\tt 1102.2219}}.
37 pages. References added, corrected typos.

\bibitem{Boulanger:2012bj}
N.~Boulanger, N.~Colombo, and P.~Sundell, ``{A Minimal BV Action for Vasiliev's
  Four Dimensional Higher Spin Gravity},''
\href{http://arXiv.org/abs/1205.3339}{{\tt 1205.3339}}.

\bibitem{Iazeolla:2007wt}
C.~Iazeolla, E.~Sezgin, and P.~Sundell, ``{Real Forms of Complex Higher Spin
  Field Equations and New Exact Solutions},'' {\em Nucl.Phys.} {\bf B791}
  (2008) 231--264,
\href{http://arXiv.org/abs/0706.2983}{{\tt 0706.2983}}.

\bibitem{Iazeolla:2008ix}
C.~Iazeolla and P.~Sundell, ``{A Fiber Approach to Harmonic Analysis of
  Unfolded Higher-Spin Field Equations},'' {\em JHEP} {\bf 0810} (2008) 022,
\href{http://arXiv.org/abs/0806.1942}{{\tt 0806.1942}}.

\bibitem{Boulanger:2008up}
N.~Boulanger, C.~Iazeolla, and P.~Sundell, ``{Unfolding Mixed-Symmetry Fields
  in AdS and the BMV Conjecture: I. General Formalism},'' {\em JHEP} {\bf 0907}
  (2009) 013,
\href{http://arXiv.org/abs/0812.3615}{{\tt 0812.3615}}.

\bibitem{Boulanger:2008kw}
N.~Boulanger, C.~Iazeolla, and P.~Sundell, ``{Unfolding Mixed-Symmetry Fields
  in AdS and the BMV Conjecture. II. Oscillator Realization},'' {\em JHEP} {\bf
  0907} (2009) 014,
\href{http://arXiv.org/abs/0812.4438}{{\tt 0812.4438}}.

\bibitem{Iazeolla:2011cb}
C.~Iazeolla and P.~Sundell, ``{Families of Exact Solutions to Vasiliev's 4D
  Equations With Spherical, Cylindrical and Biaxial Symmetry},'' {\em JHEP}
  {\bf 1112} (2011) 084,
\href{http://arXiv.org/abs/1107.1217}{{\tt 1107.1217}}.

\bibitem{Pilch:1984aw}
K.~Pilch, P.~van Nieuwenhuizen, and M.~Sohnius, ``{De Sitter Superalgebras and
  Supergravity},'' {\em Commun.Math.Phys.} {\bf 98} (1985)
105.

\bibitem{Anninos:2011ui}
D.~Anninos, T.~Hartman, and A.~Strominger, ``{Higher Spin Realization of the
  dS/CFT Correspondence},''
\href{http://arXiv.org/abs/1108.5735}{{\tt 1108.5735}}.

\bibitem{Ng:2012xp}
G.~S. Ng and A.~Strominger, ``{State/Operator Correspondence in Higher-Spin
  dS/CFT},''
\href{http://arXiv.org/abs/1204.1057}{{\tt 1204.1057}}.

\bibitem{Martelli:2008rt}
D.~Martelli and J.~Sparks, ``{Notes onToric Sasaki-Einstein Seven-manifolds and
  AdS(4)/CFT(3)},'' {\em JHEP} {\bf 0811} (2008) 016,
\href{http://arXiv.org/abs/0808.0904}{{\tt 0808.0904}}.

\bibitem{Martelli:2008si}
D.~Martelli and J.~Sparks, ``{Moduli Spaces of Chern-Simons Quiver Gauge
  Theories and AdS(4)/CFT(3)},'' {\em Phys.Rev.} {\bf D78} (2008) 126005,
\href{http://arXiv.org/abs/0808.0912}{{\tt 0808.0912}}.

\bibitem{Engquist:2002gy}
J.~Engquist, E.~Sezgin, and P.~Sundell, ``{Superspace Formulation of 4D Higher
  Spin Gauge Theory},'' {\em Nucl.Phys.} {\bf B664} (2003) 439--456,
\href{http://arXiv.org/abs/hep-th/0211113}{{\tt hep-th/0211113}}.

\bibitem{Giombi:2011kc}
S.~Giombi, S.~Minwalla, S.~Prakash, S.~P. Trivedi, S.~R. Wadia, and Y.~Shuichi,
  ``{Chern-Simons Theory with Vector Fermion Matter},''
\href{http://arXiv.org/abs/1110.4386}{{\tt 1110.4386}}.

\bibitem{Aharony:2011jz}
O.~Aharony, G.~Gur-Ari, and R.~Yacoby, ``{d=3 Bosonic Vector Models Coupled to
  Chern-Simons Gauge Theories},'' {\em JHEP} {\bf 1203} (2012) 037,
\href{http://arXiv.org/abs/1110.4382}{{\tt 1110.4382}}.

\bibitem{Giombi:2012ms}
S.~Giombi and X.~Yin, ``{The Higher Spin/Vector Model Duality},''
\href{http://arXiv.org/abs/1208.4036}{{\tt 1208.4036}}.

\bibitem{Sezgin:2005pv}
E.~Sezgin and P.~Sundell, ``{An Exact Solution of 4D Higher Spin Gauge
  Theory},'' {\em Nucl.Phys.} {\bf B762} (2007) 1--37,
\href{http://arXiv.org/abs/hep-th/0508158}{{\tt hep-th/0508158}}.

\bibitem{Colombo:2012jx}
N.~Colombo and P.~Sundell, ``{Higher Spin Gravity Amplitudes From Zero-form
  Charges},''
\href{http://arXiv.org/abs/1208.3880}{{\tt 1208.3880}}.

\bibitem{Girardello:2002pp}
L.~Girardello, M.~Porrati, and A.~Zaffaroni, ``{3D Interacting CFTs and
  Generalized Higgs Phenomenon in Higher Spin Theories on AdS},'' {\em
  Phys.Lett.} {\bf B561} (2003) 289--293,
\href{http://arXiv.org/abs/hep-th/0212181}{{\tt hep-th/0212181}}.

\end{thebibliography}


\end{document}